\documentclass[a4paper]{article}
\usepackage{a4wide}
\usepackage{graphicx}
\usepackage[T1]{fontenc}
\usepackage[utf8]{inputenc}
\usepackage[UKenglish]{babel}
\usepackage{placeins}
\usepackage[shortlabels]{enumitem}
\usepackage[binary-units=true]{siunitx}
\usepackage{color}
\usepackage{authblk}
\usepackage{amsmath}
\usepackage{amssymb}
\usepackage{amsthm}
\usepackage{mathtools}
\usepackage{bbold}
\usepackage{braket}
\usepackage[ruled,lined]{algorithm2e}
\usepackage{nicefrac}
\usepackage[backend=bibtex,style=phys,sorting=none,citestyle=numeric-comp,eprint=true]{biblatex}
\usepackage[
colorlinks=true,
allcolors=blue
]{hyperref}
\usepackage{cleveref}

\graphicspath{{figures/},{data/}}

\crefformat{footnote}{#2\footnotemark[#1]#3}

\DeclareMathOperator{\im}{i}
\DeclareMathOperator{\real}{Re}
\renewcommand{\Re}{\real}

\DeclareMathOperator{\tr}{Tr}
\DeclareMathOperator{\diag}{diag}

\DeclareMathOperator{\ord}{\mathcal{O}}
\newcommand{\pdagger}{{\phantom{\dagger}}}

\newcommand{\del}[2]{\ensuremath{\frac{\partial #1}{\partial#2}}}
\newcommand{\eto}[1]{\ensuremath{\mathrm{e}^{#1}}}
\newcommand{\trans}{\ensuremath{\mathsf{T}}}
\newcommand{\md}{\ensuremath{\mathrm{d}}}

\newcommand{\id}{\ensuremath{\mathbb{1}}}
\newcommand{\tint}{\ensuremath{\tau_\text{int}}}
\newcommand{\ordnung}[1]{\ensuremath{\ord\left(#1\right)}}
\newcommand{\erwartung}[1]{\ensuremath{\left\langle#1\right\rangle}}

\newtheorem{theorem}{Theorem}
\newtheorem{corollary}{Corollary}

\theoremstyle{definition}

\theoremstyle{remark}
\newtheorem*{remark}{Remark}

\newcommand{\bonn}{
	\textit{\footnotesize Helmholtz-Institut f\"{u}r Strahlen- und
		Kernphysik,
		University of Bonn, 53115 Bonn, Germany}
}

\newcommand{\liverpool}{
	\textit{\footnotesize Department of Mathematical Sciences,
		University of Liverpool, Liverpool, L69 7ZL, United Kingdom}
}

\makeatletter%
\makeatletter%
\begin{document}
	
	\title{Minimal Autocorrelation in Hybrid Monte Carlo simulations using Exact Fourier Acceleration}
	
	\author[1,2]{Johann Ostmeyer\footnote{\href{mailto:ostmeyer@hiskp.uni-bonn.de}{ostmeyer@hiskp.uni-bonn.de}}}
	\author[1]{Pavel Buividovich}
	\affil[1]{\liverpool}
	\affil[2]{\bonn}
	\date{\today\vspace*{-\baselineskip}}
	\maketitle
	
	\begin{abstract}
		The hybrid Monte Carlo (HMC) algorithm is a ubiquitous method in computational physics with applications ranging from condensed matter to lattice QCD and beyond. However, HMC simulations often suffer from long autocorrelation times, severely reducing their efficiency. In this work two of the main sources of autocorrelations are identified and eliminated. The first source is the sampling of the canonical momenta from a sub-optimal normal distribution, the second is a badly chosen trajectory length. Analytic solutions to both problems are presented and implemented in the exact Fourier acceleration (EFA) method. It completely removes autocorrelations for near-harmonic potentials and consistently yields (close-to-) optimal results for numerical simulations of the Su-Schrieffer-Heeger and the Ising models as well as in lattice gauge theory, in some cases reducing the autocorrelation by multiple orders of magnitude. EFA is advantageous for and easily applicable to any HMC simulation of an action that includes a quadratic part.
	\end{abstract}

	\unitlength = 1em
	
	\section{Introduction}

Originally proposed for the simulation of lattice field theory, the hybrid Monte Carlo (HMC)~\cite{Duane1987} algorithm has found its way into almost every branch of physics. It uses global updates with a high acceptance probability which allows the HMC to probe high-dimensional phase spaces. HMC simulations are viable for practically any (quantum) system, as long as its dynamics are governed by some quantifiable real-valued action.

So why does the HMC not sample a system as simple as the harmonic oscillator without autocorrelation?

In short, it does, it just has to be tuned correctly. A somewhat longer answer will be presented throughout this work. The goal is to develop a method that allows the HMC to sample harmonic problems as efficiently as the known optimal solution, i.e.\ without autocorrelation. At the same time, this method should be generalisable to arbitrary anharmonic actions. By continuity, actions governed by a harmonic part would still be sampled with negligible autocorrelation and for strongly anharmonic systems the method would not be worse than any other.

The method that fulfils all these criteria is presented in \cref{th:opt_traj_and_kin} and called exact Fourier acceleration (EFA). It is summarised in \cref{alg:efa-hmc,alg:leap-frog,alg:efa}. The idea of Fourier acceleration (FA) reaches back to Ref.~\cite{PhysRevD.32.2736}, while a scale separation of different force contributions has been introduced in Ref.~\cite{Sexton:1992nu,URBACH200687} and used in combination with FA in Ref.~\cite{Scalettar:2203.01291}. The scale separation is required here for the individual treatment of the quadratic or harmonic part of the action. Recently, in Ref.~\cite{Ostmeyer:2023azi} we have introduced a first version of EFA. This work generalises EFA to actions with arbitrary quadratic terms, not just those diagonalisable by Fourier transformation. Moreover, the importance of the trajectory length is identified and the optimal value derived.

The rest of this work is structured as follows. In \cref{sec:raw_hmc} the HMC algorithm in its classical form is recalled, followed by the optimised formulation in \cref{sec:efa_opt}. Deviations from this optimal form are explored and their implications quantified in \cref{sec:bad_params}. A discussion of strongly anharmonic potentials (typically caused by strong interactions) concludes \cref{sec:formalism}. The entire \cref{sec:examples} lists numerical examples supporting the theoretical results from \cref{sec:formalism} and investigating non-trivial edge cases. The physical systems included are the Su-Schrieffer-Heeger model in \cref{sec:ssh}, the classical Ising model in \cref{sec:ising}, the Hubbard model in \cref{sec:hubbard_model}, and lattice gauge theory in \cref{sec:qcd}. The findings are recapitulated in \cref{sec:conclusion}. 	
	\section{Formalism and Theory}\label{sec:formalism}

\subsection{The hybrid Monte Carlo algorithm}\label{sec:raw_hmc}

The HMC algorithm is a Markov chain Monte Carlo (MCMC) method to sample continuous fields $x$ from a given probability distribution
$	x \sim \eto{-S(x)}$
defined by the action $S(x)$ using global updates~\cite{Duane1987}. An update is proposed based on the previous configuration and some random momentum $p$ sampled from a normal distribution
$	p \sim \eto{-\frac{p^2}{2m^2}}$
with some masses $m$ that will be considered in detail in \cref{sec:efa_opt}. The combined probability distribution $x,p\sim \eto{-\mathcal{H}}$ is governed by the Hamiltonian
\begin{align}
	\mathcal{H} &= \frac{p^2}{2m^2} + S(x)\,.\label{eq:simple_hmc_hamiltonian}
\end{align}
The idea is that updates with a high acceptance probability can be proposed by solving the equations of motion (EOM)
\begin{align}
	\dot x &= \phantom{-}\del{\mathcal{H}}{p}\,,\\
	\dot p &= - \del{\mathcal{H}}{x}\,,
\end{align}
which preserve the Hamiltonian, over some finite trajectory length $T$. In practice, the EOM have to be solved numerically (typically using a symplectic integrator~\cite{OMELYAN2003272}) so that $\mathcal{H}$ cannot be preserved exactly. In order to restore the correct probability distribution, the change $\Delta \mathcal{H}$ is compensated using a final accept/reject step. The acceptance probability for the new configuration $x(T)$ is given by the Boltzmann weight
\begin{align}
	p_\text{acc} &= \min\left(1, \eto{-\Delta \mathcal{H}}\right)\,.
\end{align}

The complete HMC algorithm, using the optimisations from \cref{th:opt_traj_and_kin}, is summarised in \cref{alg:efa-hmc}.

\begin{algorithm*}[tb]
	\caption{Full hybrid Monte Carlo (HMC) trajectory update with EFA (alg.~\ref{alg:efa}) given an integrator (e.g.\ leap-frog, alg.~\ref{alg:leap-frog}).}\label{alg:efa-hmc}
	\SetKwInOut{Input}{input}
	\SetKwInOut{Params}{parameters}
	\SetKwInOut{Output}{output}
	\Input{initial fields $x^\text{i}$, molecular dynamics steps $N_\text{MD}$, trajectory length $T=\frac\pi2$}
	\Params{harmonic matrix $M=\Omega\cdot\diag(\omega^2)\cdot\Omega^\dagger$, anharmonic potential $V$}
	\Output{final fields $x^\text{f}$}
	$x \gets x^\text{i}$\;
	sample $r\sim \mathcal{N}(0,1)^{\dim(M)}$ \tcp*{standard normal distribution}
	$p \gets \Omega\cdot\diag(\omega)\cdot\Omega^\dagger \cdot r$ \tcp*{any realisation of $p\gets\sqrt{M}\cdot r$ can be used}
	$\mathcal{H}^\text{i} \gets \frac12 r^2 + \frac12 x^\trans M x + V(x)$ \tcp*{use $p^\trans M^{-1} p = r^2$}
	\For{$\tau \gets 1\dots N_\text{MD}$}{
		$(x,p) \gets \text{integrator}\left(x,p,\nicefrac{T}{N_\text{MD}}\right)$\;
	}
	$\mathcal{H}^\text{f} \gets \frac12 p^\trans M^{-1} p + \frac12 x^\trans M x + V(x)$\;
	$\Delta \mathcal{H} \gets \mathcal{H}^\text{f}-\mathcal{H}^\text{i}$\;
	\uIf(\tcp*[f]{uniform distribution}){$\eto{-\Delta \mathcal{H}}\ge \mathcal{U}_{[0,1]}$}{$x^\text{f} \gets x$\;}
	\Else{$x^\text{f} \gets x^\text{i}$\;}
\end{algorithm*}

A multitude of alternative methods can be used to sample non-trivial probability distributions. The HMC is typically a good choice for large systems and, as we derive in this work, for actions with a dominating harmonic part. Some of the most prominent alternative methods (in physics) include the local updating scheme by Blankenbecler and Scalapino, and Sugar (BSS)~\cite{BSS:1981} and normalizing flow~\cite{Tabak:2010,Kobyzev_2021}. The BSS algorithm can be highly efficient for small to medium size systems, in particular close to phase transitions. Normalizing flow, on the other hand, allows to sample without autocorrelation as long as the target distribution can be approximated well by a relatively simple invertible differentiable transformation of a distribution that can be sampled directly. It shares the concept with the HMC that sampling (usually) starts out with a normal distribution that is subsequently adjusted according to the target distribution.

The most important parameter for the quantification of the sampling efficiency in HMC (or any other MCMC) simulations is the integrated autocorrelation time
\begin{align}
	\tint &\equiv \frac12 + \sum_{t=1}^{\infty} \rho_\mathcal{A}(t)\,,
\end{align}
where $\rho_\mathcal{A}(t)$ denotes the autocorrelation function of the observable $\mathcal{A}$ after HMC time $t$. In practice, the sum has to be truncated. A reliable way to do so with minimal statistical and systematic errors is described in Ref.~\cite{WOLFF2004143}. \tint\ effectively measures how long it will take the Markov chain to produce an independent configuration. That is, given a precision goal, the required compute time is simply proportional to \tint. The best possible sampling decorrelates successive configurations completely and thus results in the lowest possible value of $\tint=\frac12$.

\subsection{Exact Fourier acceleration and optimal HMC parameters}\label{sec:efa_opt}

To date, there is no known way to derive the set of HMC parameters analytically that minimises \tint\ for arbitrary actions. The best one can hope for is to eliminate autocorrelations due to analytically tractable (harmonic) parts of the action. More specifically, we define an HMC parameter set as optimal if it guarantees full decorrelation of the field $x$ after a single trajectory in the absence of an anharmonic potential $V(x)$. As long as $V(x)$ does not dominate the action, such an optimal set also minimises the autocorrelation of the full dynamics. This allows us to quantify said optimal set in the following theorem.

\begin{theorem}[Optimal HMC trajectory length and kinetic term]\label{th:opt_traj_and_kin}
	Let $S$ be an action of the form
	\begin{align}
		S(x) &= \frac12 x^\trans M x + V(x)\label{eq:general_action}
	\end{align}
	with a constant hermitian positive definite matrix $M$ and an arbitrary anharmonic potential $V(x)$. Then the Hamiltonian
	\begin{align}
		\mathcal{H} &= \frac12 p^\trans  M^{-1}  p + S(x)\label{eq:best_hamiltonian}
	\end{align}
	together with the HMC trajectory length $T=\frac{\pi}{2}$ yield optimal performance in HMC simulations in the sense that they completely remove autocorrelations stemming from the harmonic part $\frac12 x^\trans M x$ of the action.
\end{theorem}

\begin{remark}
	In many practical applications $M$ is translationally invariant and can be diagonalised and inverted via Fourier transformation. The idea to use $M^{-1}$ in the kinetic term first arose in this context which is why this choice is typically called Fourier acceleration (FA). We will stick to this historically motivated nomenclature, however \cref{th:opt_traj_and_kin} is more generally applicable and ``harmonic'' acceleration might be a more appropriate name. The optimal choice of $M^{-1}$ together with $T=\frac\pi2$ has been used in the special case when $M$ can be inverted with a Fourier transformation~\cite{Borsanyi:2015,Cianci:2023vtm}. We also note that the generalised use of $M^{-1}$ has been advocated before~\cite{Riemann_Manifold:2011,Neal:2011mrf}, but to our knowledge not in conjunction with the optimal trajectory length.
\end{remark}

\begin{proof}
	The anharmonic potential $V(x)$ is ignored in the following because it is irrelevant in our notion of optimality. Choose $M=\diag(\omega^2)$ diagonal ($M$ is diagonalisable because it is hermitian and all eigenvalues $\omega_i^2$, $i=1,\dots,\dim(M)$ are demanded to be strictly positive) and set the kinetic term from equation~\eqref{eq:simple_hmc_hamiltonian} to $\frac12 \sum_i p_i^2/m_i^2$, explicitly allowing different `masses' $m_i$ for each component. The equations of motion (EOM) then read
	\begin{align}
		\dot x_i &= \frac{p_i}{m_i^2}\,,\\
		\dot p_i &= -\omega_i^2 x_i\,.
	\end{align}
	These harmonic EOM are solved by
	\begin{align}
		x_i(t) &= x_i^0 \cos\left(\frac{\omega_i}{m_i}t\right) + \frac{1}{m_i\omega_i}\, p_i^0 \sin\left(\frac{\omega_i}{m_i}t\right)\,,\label{eq:eom_sol_x_of_t}\\
		p_i(t) &= p_i^0 \cos\left(\frac{\omega_i}{m_i}t\right) - m_i\omega_i\, x_i^0 \sin\left(\frac{\omega_i}{m_i}t\right)\,.
	\end{align}
	Now the optimal choice of the `masses' $m_i$ and the trajectory length $T$ is such that $x(T)$ is completely decorrelated from $x(0)$. This is the case if for all $i$
	\begin{align}
		\cos\left(\frac{\omega_i}{m_i}T\right) &= 0\\
		\Leftrightarrow \frac{\omega_i}{m_i}T &= \frac\pi2 \pmod \pi\,.
	\end{align}
	There are infinitely many solutions to this condition (for instance, every value of $T\neq0$ is allowed), but the most natural choice is
	\begin{align}
		m_i &= \omega_i\,,\label{eq:m_i_eq_omega_i}\\
		T &= \frac{\pi}{2}\,.\label{eq:T_is_pi2}
	\end{align}
	Since the kinetic term had been chosen diagonal in the same basis as $M$ and all its eigenvalues are inverse to those of $M$, we deduce that $\frac12 p^\trans M^{-1} p$ is an optimal kinetic term in every basis.
\end{proof}

This proof is highly instructive in that it provides us with an analytic exact solution to the harmonic EOM. This solution can be used to augment the numerical integration of the EOM by solving the harmonic (or Fourier) part exactly and only treating the anharmonic forces $-\nabla V(x)$ numerically, see \cref{alg:leap-frog,alg:efa}. As in Ref.~\cite{Ostmeyer:2023azi}, this method is called exact Fourier acceleration (EFA). Note that the initial formulation of EFA~\cite{Ostmeyer:2023azi} did not include the optimal parameter choice as in \cref{eq:m_i_eq_omega_i,eq:T_is_pi2}. It should not be used since the version introduced in this work is strictly superior. From now on we will use the term EFA exclusively to denote the optimal algorithm introduced here.

\begin{algorithm*}[tb]
	\caption{Single update step with the leap-frog integrator~\cite{Verlet:1967,trotter_omelyan} and EFA (alg.~\ref{alg:efa}).}\label{alg:leap-frog}
	\SetKwInOut{Input}{input}
	\SetKwInOut{Params}{parameters}
	\SetKwInOut{Output}{output}
	\Input{initial fields $x^0$, momenta $p^0$, time step $h$}
	\Params{anharmonic forces $-\nabla V$}
	\Output{final fields $x(h)$ and momenta $p(h)$}
	$(x,p) \gets \text{EFA}\left(x^0,p^0,\nicefrac h2\right)$\;
	$p \gets p - h\cdot\nabla V(x)$\;
	$\left(x(h),p(h)\right) \gets \text{EFA}\left(x,p,\nicefrac h2\right)$\;
\end{algorithm*}

\begin{algorithm*}[tb]
	\caption{Single time step using exact Fourier acceleration (EFA).}\label{alg:efa}
	\SetKwInOut{Input}{input}
	\SetKwInOut{Params}{parameters}
	\SetKwInOut{Output}{output}
	\Input{initial fields $x^0$, momenta $p^0$, time step $h$}
	\Params{harmonic matrix $M=\Omega\cdot\diag(\omega^2)\cdot\Omega^\dagger$}
	\Output{final fields $x(h)$ and momenta $p(h)$}
	$y^0 \gets \Omega^\dagger \cdot x^0$ \tcp*{$\Omega$ is often a Fourier transformation, thence the name EFA}
	$q^0 \gets \Omega^\dagger \cdot p^0$\;
	\For{$i \gets 1\dots \dim(M)$}{
		$y_i(h) \gets \cos(h)\, y^0_i + \frac{1}{\omega_i^2} \sin(h)\, q^0_i$\;
		$q_i(h) \gets \cos(h)\, q^0_i - \omega_i^2 \sin(h)\, y^0_i$\;
	}
	$x(h) \gets \Omega \cdot y(h)$\;
	$p(h) \gets \Omega \cdot q(h)$\;
\end{algorithm*}

In a direct comparison of classical FA~\cite{PhysRevD.32.2736,Scalettar:2203.01291,PhysRevD.56.6885,Catterall:2001jg,Flamino:2018jmo,Horsley:2023fhj} (i.e.\ using a numerical solution to the EOM) and EFA, one finds that
EFA has a number of advantages and no disadvantages. By far the most important advantage is that it allowed for the choice of the optimal trajectory length in \cref{th:opt_traj_and_kin}. It also comes with a computational speed up since the harmonic part of the trajectory does not have to be divided into discrete steps, each requiring additional FFTs. In practice this speed up is often negligible compared to the cost of calculating the anharmonic potential, typically involving fermionic contributions with determinants or inversions. For practical applications it turns out to be more useful that EFA comes with the smallest possible energy violation and thus highest possible acceptance rate without the need to tune the number of discrete update steps.

Overall, clearly EFA should be used instead of classical FA whenever possible, however the main contribution to the acceleration comes from the use of \cref{th:opt_traj_and_kin} and not from the exact solution of the harmonic EOM.

\subsection{Implication of sub-optimal parameters}\label{sec:bad_params}

In order to understand the significance of \cref{th:opt_traj_and_kin}, let us explore the consequences of other parameter choices quantitatively.

The proofs of both \cref{th:no_fourier_acc,th:tau_int_short_traj} proceed by direct calculation and can be found in \cref{sec:proofs}.

\begin{corollary}[No Fourier acceleration]\label{th:no_fourier_acc}
	If no FA is used, i.e.\ instead of the Hamiltonian~\eqref{eq:best_hamiltonian} one chooses $\mathcal{H} = \frac 12 p^2 + S(x)$ with fixed trajectory length, then the integrated autocorrelation time $\tau_\mathrm{int}$ of an observable $\mathcal{A}$ will be rescaled by the factor
	\begin{align}
		\tau_\mathrm{int} & \propto \left(\frac{\omega_\mathrm{max}}{\omega_\mathrm{min}}\right)^2 + \ordnung{1}\,,\label{eq:no_fa_rescaling}
	\end{align}
	where $\omega_\mathrm{min}^2$ denotes the smallest eigenvalue of $M$ with non-zero overlap of the corresponding eigenstate $v_\text{min}$ with the observable $\Braket{v_\mathrm{min}|\mathcal{A}|v_\mathrm{min}}\neq 0$ and $\omega_\mathrm{max}^2$ is the biggest eigenvalue of $M$.
\end{corollary}

	This scenario is very common and explains why most HMC simulations suffer from high autocorrelation times. For practical purposes one can typically assume that $\omega_\text{min}^2$ is simply the smallest eigenvalue of $M$, unless one can show that the corresponding eigenstate has no overlap with the operator $\mathcal{A}$. Thus, the rescaling factor is the condition number of $M$ in most cases.

The result of \cref{th:no_fourier_acc} is also found in~\cite{KENNEDY1991118}, though a slightly different derivation and a less general setting are used. Recently, in Ref.~\cite{apers2022hamiltonian} it has been shown that the total computational cost of HMC sampling (up to logarithmic corrections) can be reduced to $\ordnung{\frac{\omega_\mathrm{max}}{\omega_\mathrm{min}}}$ even without FA by ensuring a long enough trajectory length $T_\text{max} \gtrsim \omega_\mathrm{min}^{-1}$ and sampling $T\in(0,T_\text{max}]$ randomly. In this scenario $\tint$ remains constant and the cost of a single trajectory scales with $\omega_\mathrm{max}/\omega_\mathrm{min}$.

\begin{corollary}[Too short trajectory]\label{th:tau_int_short_traj}
	Presuming FA as in equation~\eqref{eq:best_hamiltonian} is used together with some short trajectory length $T<\frac\pi2$ and measurements of an observable are conducted in intervals of fixed HMC time, the integrated autocorrelation time of this observable is given by
	\begin{align}
		\tau_\mathrm{int}(T) &= \frac{\tau_\mathrm{int}(T=\nicefrac{\pi}{2})}{ 1-\cos(T)^{\nicefrac{\pi}{2T}}}\,.\label{eq:short_traj_rescaling}
	\end{align}
\end{corollary}

	Equation~\eqref{eq:short_traj_rescaling} can be expanded in short trajectory lengths $T$ yielding
	\begin{align}
		\left(1-\cos(T)^{\nicefrac{\pi}{2T}}\right)^{-1} &= \frac{4}{\pi T} + \frac12 + \ordnung{T}\,.
	\end{align}
	Thus, choosing a short trajectory length leads to a linear divergence of the autocorrelation time in $1/T$, while neglecting to use FA at fixed $T$ results in quadratic divergence in the frequency ratio $\omega_\mathrm{max}/\omega_\mathrm{min}$ or, equivalently, linear divergence in the condition number of $M$.
	Following Ref.~\cite{apers2022hamiltonian}, linear divergence of the cost in $\omega_\mathrm{max}/\omega_\mathrm{min}$ can be achieved without FA using correspondingly long trajectories.

\Cref{th:tau_int_short_traj} is only valid for $T<\nicefrac\pi2$ and equation~\eqref{eq:short_traj_rescaling} breaks down for $T>\nicefrac\pi2$ because of the non-integer power of the negative $\cos(T)$. Nevertheless it makes sense to consider the effect of longer trajectories. In the extreme case of $T=\pi$ every new configuration is simply given by $x(T)=-x(0)$, resulting in maximal anti-correlation. This non-ergodic regime is as useless as maximal correlation $x(T)=x(0)$ achieved by $T=0\pmod{2\pi}$, but it is more deceptive because a naively calculated integrated autocorrelation time would be very low. In some strongly anharmonic cases (see sec.~\ref{sec:ising}) it can appear advantageous to choose $T>\nicefrac\pi2$, but this has to be done with the utmost care, verifying that the resulting configurations are truly decorrelated and not anti-correlated.

\subsection{Deviations for strongly anharmonic potentials}\label{sec:anharmonic}

For the purely harmonic case, i.e.\ $V(x)=0$ in equation~\eqref{eq:general_action}, \cref{th:opt_traj_and_kin} allows to sample from the normal distribution $x\sim\exp(-\nicefrac12\, x^\trans M x)$ in the HMC framework without any autocorrelation whatsoever. By continuity, autocorrelation will remain small for small anharmonic potentials $|V(x)|\ll x^\trans M x$. However, there is no guarantee that \cref{th:opt_traj_and_kin} still provides the best sampling method for strongly anharmonic actions $|V(x)| \gtrsim \nicefrac12\, x^\trans M x$. As a rule, it is still a good starting point, but some amendments can be necessary.

For strongly anharmonic potentials it is not known in general whether they allow periodic behaviour and, if so, with which periodicity. It is possible, even likely, that no periodicities (close to the chosen trajectory length) occur. However, this cannot be guaranteed and a fixed trajectory length $T$ can lead to (almost) periodic trajectories, increasing autocorrelation or even inhibiting ergodicity, as has been observed in Ref.~\cite{cohenstead2024smoqydqmcjl} (using an augmented version of EFA from~\cite{Ostmeyer:2023azi}).

This problem can easily be avoided by choosing a variable random trajectory length~\cite{Mackenzie:1989us}. No-U-turn sampling~\cite{NUTS-2014} is an example of a very elaborate algorithm that effectively chooses a random $T\in[-\pi,\pi]$ without assuming any prior knowledge about the periodicity. A simple and yet effective alternative method is to choose $T$ uniformly from a fixed interval as in Ref.~\cite{cohenstead2024smoqydqmcjl}. The theoretically optimal interval is $T\in\left[\nicefrac\pi2-\delta,\nicefrac\pi2+\delta\right]$ with $0\le\delta\le\nicefrac\pi2$. For small anharmonic potentials $V(x)$ one can safely choose $\delta=0$ and only for very large $V(x)$ some $\delta>0$ is required. The precise value has to be adjusted during numerical experiments and depends on the specifics of the system. In all simulations presented in this work, a constant trajectory length was found sufficient to obtain unbiased results and this direction has not been further explored.

As a rule, classical FA as in~\cite{Scalettar:2203.01291} is used with a regulator $m_\text{reg}>0$ choosing $m_i=\sqrt{m_\text{reg}^2+\omega_i^2}$ instead of the exact formula~\eqref{eq:m_i_eq_omega_i}. There is no need for such a regulator in simulations of close to harmonic actions with EFA. In fact, a bad choice of $m_\text{reg}$ can restore a situation closer to that without FA and thus considerably slow down simulations (see \cref{th:no_fourier_acc}). However, a finite $m_\text{reg}$ can be advantageous in strongly anharmonic cases~\cite{cohenstead2024smoqydqmcjl}.

In some simulations numerical stability plays an important role. Since numerical errors accumulate over the trajectory, these simulations can require a shorter trajectory length $T<\nicefrac\pi2$ than in \cref{th:opt_traj_and_kin}. This behaviour is discussed in detail \cref{sec:ssh}.

Notably, FA cannot overcome critical slowing down caused by anharmonic effects. More specifically, FA is designed to maximise sampling efficiency within a potential minimum (locally described by a parabola), but it does not provide a means to tunnel from one local minimum to another. The example in \cref{sec:ising} vividly demonstrates this limitation.

Finally, while \cref{th:opt_traj_and_kin} guarantees (instant) thermalisation for harmonically dominated actions, no such guarantee is given for general actions. In particular, on non-compact manifolds the HMC (independently of FA) needs to be augmented by a so-called radial update in order to guarantee thermalisation~\cite{original_radial_update,Ostmeyer:2024gnh}. Such a radial update introduces non-local steps, e.g.\ multiplicative updates of the current state, and thus allows to overcome slow diffusive behaviour in remote regions of the phase space. It can also restore ergodicity to the simulations of some actions with phase space separation~\cite{Temmen:2024pcm,Temmen_ergodicity}. 	
	\section{Applications and numerical examples}\label{sec:examples}
	
	In the following the implications of \cref{th:opt_traj_and_kin} as well as \cref{th:no_fourier_acc,th:tau_int_short_traj} will be explored using a variety of numerical examples.
	To start with, the efficacy of EFA as in \cref{th:opt_traj_and_kin} is demonstrated in \cref{sec:ssh} and the proportionality predicted by \cref{th:no_fourier_acc} is verified.
	\Cref{sec:ising} further elaborates on the importance to choose the optimal trajectory length in accordance with \cref{th:tau_int_short_traj}.
	This section concludes with a method that allows to apply FA to lattice gauge theories in \cref{sec:qcd}.
	An example of purely ``harmonic'' as opposed to Fourier acceleration is provided in \cref{sec:hubbard_model}.
	
	The integrated autocorrelation times $\tau_\text{int}$ from numerical results presented in this work have been calculated following the prescription of Ref.~\cite{WOLFF2004143}.
	
	\subsection{The Su-Schrieffer-Heeger model}\label{sec:ssh}

Su-Schrieffer-Heeger (SSH)~\cite{PhysRevLett.42.1698} type models are well known for their notoriously bad autocorrelation times~\cite{PhysRevLett.126.017601,Assaad:2102.08899, Cai:2308.06222}.
FA has been found advantageous in their simulations before~\cite{PhysRevB.99.035114,Ostmeyer:2023azi} and it turns out that EFA as in \cref{th:opt_traj_and_kin} can almost completely remove autocorrelation as will be demonstrated in the following.

The version of the SSH model used here describes spinless electrons with phonon-mediated interactions in two dimensions. It is defined by the Hamiltonian
\begin{align}
	H_\text{SSH} = 
	\omega_0\sum_{i,\alpha}\left(a^\dagger_{i,\alpha}a^\pdagger_{i,\alpha} + \frac12\right)
	- \sum_{i,\alpha} {J}_\alpha \left(1-\lambda_\alpha x_{i,\alpha}\right) \left(c_i^\dagger c^\pdagger_{i+\alpha} + c^\dagger_{i+\alpha}c^\pdagger_{i} \right)
	- \mu \sum_{i} c_i^\dagger c^\pdagger_i\,,
	\label{eq:hamiltonian_ssh}
\end{align}
where the creation (annihilation) operators $a^\dagger_{i,\alpha}$ ($a^\pdagger_{i,\alpha}$) and associated fields $x_{i,\alpha}\equiv (a^\dagger_{i,\alpha} + a^\pdagger_{i,\alpha})/\sqrt{2\omega_0}$ describe the phonons on the $\alpha$-th link of site $i$. Their dynamics are governed by a harmonic oscillator with frequency $\omega_0$. The nearest neighbour hopping amplitudes $J_\alpha$ of the electrons $c^\dagger_i$, $c^\pdagger_i$ along the link $\alpha$ are modulated by the phonon fields $x_{i,\alpha}$ with the coupling strength $\lambda_\alpha$. A chemical potential $\mu$ is applied.

In Ref.~\cite{Ostmeyer:2023azi} the path integral formulation of this model at inverse temperature $\beta$ is derived, resulting in an action of the form
\begin{align}
	S_\text{SSH} &= \frac{1}{2} x^\trans\, M_\text{SSH}\, x \,+\, \text{electron interactions}\\
	&= \frac{\beta}{2N_t} \sum_t \left[\omega_0^2 x_t^2 + \frac{N_t^2}{\beta^2}\left(x_{t+1}-x_t\right)^2\right] \,+\, \text{electron interactions}\\
	&= \frac{1}{2\beta N_t} \sum_\xi \left[\left(\beta\omega_0\right)^2 + 4N_t^2\sin^2\left(\frac{\pi}{N_t}\xi\right)\right] y_\xi^2 \,+\, \text{electron interactions}\,,\\
	y_{\xi} &= \frac{1}{\sqrt{N_t}}\sum_t \eto{-\im \frac{2\pi}{N_t} \xi t}\, x_{t}\,,\quad \xi=0,\dots,N_t-1\,.	
\end{align}
In order to approach the continuum limit, Euclidean time has to be discretised into a high number of time slices $N_t\gg\beta\omega_0$. Thus, the eigenvalues $\left(\beta\omega_0\right)^2 + 4N_t^2\sin^2\left(\frac{\pi}{N_t}\xi\right)$ of the harmonic matrix $M_\text{SSH}$ can span many orders of magnitude. This immediately explains the long autocorrelations in HMC simulations without FA. On the other hand, since $M_\text{SSH}$ is diagonalised exactly by Fourier transformation, this is a prototypical example for the application of EFA following \cref{th:opt_traj_and_kin}.

Two very different regimes are explored in the following. First, the case of very low filling inducing very weak interactions is considered. This case offers a realistic description of organic molecular semiconductors~\cite{Ostmeyer:2023azi,TroisiNatureMaterials2017,FratiniNikolkaNatMat2020}. The second, strongly interacting regime is more interesting from a theoretical point of view as it is in a spontaneously broken phase with charge density wave (CDW) order~\cite{PhysRevLett.126.017601}.

\begin{figure*}[t]
	\centering
	\resizebox{0.98\textwidth}{!}{{\large%
\begingroup
  \inputencoding{latin1}%
  \makeatletter
  \providecommand\color[2][]{%
    \GenericError{(gnuplot) \space\space\space\@spaces}{%
      Package color not loaded in conjunction with
      terminal option `colourtext'%
    }{See the gnuplot documentation for explanation.%
    }{Either use 'blacktext' in gnuplot or load the package
      color.sty in LaTeX.}%
    \renewcommand\color[2][]{}%
  }%
  \providecommand\includegraphics[2][]{%
    \GenericError{(gnuplot) \space\space\space\@spaces}{%
      Package graphicx or graphics not loaded%
    }{See the gnuplot documentation for explanation.%
    }{The gnuplot epslatex terminal needs graphicx.sty or graphics.sty.}%
    \renewcommand\includegraphics[2][]{}%
  }%
  \providecommand\rotatebox[2]{#2}%
  \@ifundefined{ifGPcolor}{%
    \newif\ifGPcolor
    \GPcolortrue
  }{}%
  \@ifundefined{ifGPblacktext}{%
    \newif\ifGPblacktext
    \GPblacktexttrue
  }{}%
  \let\gplgaddtomacro\g@addto@macro
  \gdef\gplbacktext{}%
  \gdef\gplfronttext{}%
  \makeatother
  \ifGPblacktext
    \def\colorrgb#1{}%
    \def\colorgray#1{}%
  \else
    \ifGPcolor
      \def\colorrgb#1{\color[rgb]{#1}}%
      \def\colorgray#1{\color[gray]{#1}}%
      \expandafter\def\csname LTw\endcsname{\color{white}}%
      \expandafter\def\csname LTb\endcsname{\color{black}}%
      \expandafter\def\csname LTa\endcsname{\color{black}}%
      \expandafter\def\csname LT0\endcsname{\color[rgb]{1,0,0}}%
      \expandafter\def\csname LT1\endcsname{\color[rgb]{0,1,0}}%
      \expandafter\def\csname LT2\endcsname{\color[rgb]{0,0,1}}%
      \expandafter\def\csname LT3\endcsname{\color[rgb]{1,0,1}}%
      \expandafter\def\csname LT4\endcsname{\color[rgb]{0,1,1}}%
      \expandafter\def\csname LT5\endcsname{\color[rgb]{1,1,0}}%
      \expandafter\def\csname LT6\endcsname{\color[rgb]{0,0,0}}%
      \expandafter\def\csname LT7\endcsname{\color[rgb]{1,0.3,0}}%
      \expandafter\def\csname LT8\endcsname{\color[rgb]{0.5,0.5,0.5}}%
    \else
      \def\colorrgb#1{\color{black}}%
      \def\colorgray#1{\color[gray]{#1}}%
      \expandafter\def\csname LTw\endcsname{\color{white}}%
      \expandafter\def\csname LTb\endcsname{\color{black}}%
      \expandafter\def\csname LTa\endcsname{\color{black}}%
      \expandafter\def\csname LT0\endcsname{\color{black}}%
      \expandafter\def\csname LT1\endcsname{\color{black}}%
      \expandafter\def\csname LT2\endcsname{\color{black}}%
      \expandafter\def\csname LT3\endcsname{\color{black}}%
      \expandafter\def\csname LT4\endcsname{\color{black}}%
      \expandafter\def\csname LT5\endcsname{\color{black}}%
      \expandafter\def\csname LT6\endcsname{\color{black}}%
      \expandafter\def\csname LT7\endcsname{\color{black}}%
      \expandafter\def\csname LT8\endcsname{\color{black}}%
    \fi
  \fi
    \setlength{\unitlength}{0.0500bp}%
    \ifx\gptboxheight\undefined%
      \newlength{\gptboxheight}%
      \newlength{\gptboxwidth}%
      \newsavebox{\gptboxtext}%
    \fi%
    \setlength{\fboxrule}{0.5pt}%
    \setlength{\fboxsep}{1pt}%
\begin{picture}(7200.00,5040.00)%
    \gplgaddtomacro\gplbacktext{%
      \csname LTb\endcsname%
      \put(1078,1076){\makebox(0,0)[r]{\strut{}$1$}}%
      \csname LTb\endcsname%
      \put(1078,2012){\makebox(0,0)[r]{\strut{}$10$}}%
      \csname LTb\endcsname%
      \put(1078,2948){\makebox(0,0)[r]{\strut{}$100$}}%
      \csname LTb\endcsname%
      \put(1078,3883){\makebox(0,0)[r]{\strut{}$1000$}}%
      \csname LTb\endcsname%
      \put(1078,4819){\makebox(0,0)[r]{\strut{}$10000$}}%
      \csname LTb\endcsname%
      \put(1210,484){\makebox(0,0){\strut{}$1$}}%
      \csname LTb\endcsname%
      \put(3074,484){\makebox(0,0){\strut{}$10$}}%
      \csname LTb\endcsname%
      \put(4939,484){\makebox(0,0){\strut{}$100$}}%
      \csname LTb\endcsname%
      \put(6803,484){\makebox(0,0){\strut{}$1000$}}%
    }%
    \gplgaddtomacro\gplfronttext{%
      \csname LTb\endcsname%
      \put(209,2761){\rotatebox{-270}{\makebox(0,0){\strut{}$\tau_\text{int}$}}}%
      \put(4006,154){\makebox(0,0){\strut{}$\omega_0 [\si{\milli\eV}]$}}%
      \csname LTb\endcsname%
      \put(2926,4591){\makebox(0,0)[r]{\strut{}EFA, $\erwartung{n_\text{el}}$}}%
      \csname LTb\endcsname%
      \put(2926,4261){\makebox(0,0)[r]{\strut{}no FA, $\erwartung{n_\text{el}}$}}%
      \csname LTb\endcsname%
      \put(2926,3931){\makebox(0,0)[r]{\strut{}prediction}}%
      \csname LTb\endcsname%
      \put(2926,3601){\makebox(0,0)[r]{\strut{}no FA, $\erwartung{n_\text{ph}}$}}%
    }%
    \gplbacktext
    \put(0,0){\includegraphics{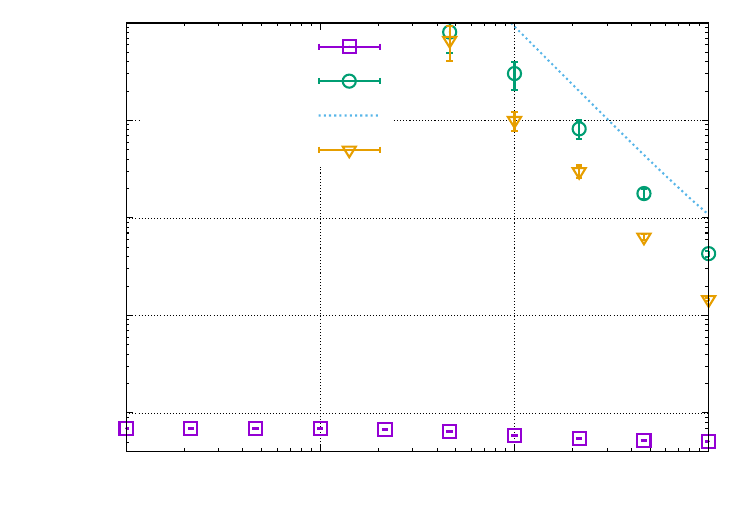}}%
    \gplfronttext
  \end{picture}%
\endgroup
\begingroup
  \inputencoding{latin1}%
  \makeatletter
  \providecommand\color[2][]{%
    \GenericError{(gnuplot) \space\space\space\@spaces}{%
      Package color not loaded in conjunction with
      terminal option `colourtext'%
    }{See the gnuplot documentation for explanation.%
    }{Either use 'blacktext' in gnuplot or load the package
      color.sty in LaTeX.}%
    \renewcommand\color[2][]{}%
  }%
  \providecommand\includegraphics[2][]{%
    \GenericError{(gnuplot) \space\space\space\@spaces}{%
      Package graphicx or graphics not loaded%
    }{See the gnuplot documentation for explanation.%
    }{The gnuplot epslatex terminal needs graphicx.sty or graphics.sty.}%
    \renewcommand\includegraphics[2][]{}%
  }%
  \providecommand\rotatebox[2]{#2}%
  \@ifundefined{ifGPcolor}{%
    \newif\ifGPcolor
    \GPcolortrue
  }{}%
  \@ifundefined{ifGPblacktext}{%
    \newif\ifGPblacktext
    \GPblacktexttrue
  }{}%
  \let\gplgaddtomacro\g@addto@macro
  \gdef\gplbacktext{}%
  \gdef\gplfronttext{}%
  \makeatother
  \ifGPblacktext
    \def\colorrgb#1{}%
    \def\colorgray#1{}%
  \else
    \ifGPcolor
      \def\colorrgb#1{\color[rgb]{#1}}%
      \def\colorgray#1{\color[gray]{#1}}%
      \expandafter\def\csname LTw\endcsname{\color{white}}%
      \expandafter\def\csname LTb\endcsname{\color{black}}%
      \expandafter\def\csname LTa\endcsname{\color{black}}%
      \expandafter\def\csname LT0\endcsname{\color[rgb]{1,0,0}}%
      \expandafter\def\csname LT1\endcsname{\color[rgb]{0,1,0}}%
      \expandafter\def\csname LT2\endcsname{\color[rgb]{0,0,1}}%
      \expandafter\def\csname LT3\endcsname{\color[rgb]{1,0,1}}%
      \expandafter\def\csname LT4\endcsname{\color[rgb]{0,1,1}}%
      \expandafter\def\csname LT5\endcsname{\color[rgb]{1,1,0}}%
      \expandafter\def\csname LT6\endcsname{\color[rgb]{0,0,0}}%
      \expandafter\def\csname LT7\endcsname{\color[rgb]{1,0.3,0}}%
      \expandafter\def\csname LT8\endcsname{\color[rgb]{0.5,0.5,0.5}}%
    \else
      \def\colorrgb#1{\color{black}}%
      \def\colorgray#1{\color[gray]{#1}}%
      \expandafter\def\csname LTw\endcsname{\color{white}}%
      \expandafter\def\csname LTb\endcsname{\color{black}}%
      \expandafter\def\csname LTa\endcsname{\color{black}}%
      \expandafter\def\csname LT0\endcsname{\color{black}}%
      \expandafter\def\csname LT1\endcsname{\color{black}}%
      \expandafter\def\csname LT2\endcsname{\color{black}}%
      \expandafter\def\csname LT3\endcsname{\color{black}}%
      \expandafter\def\csname LT4\endcsname{\color{black}}%
      \expandafter\def\csname LT5\endcsname{\color{black}}%
      \expandafter\def\csname LT6\endcsname{\color{black}}%
      \expandafter\def\csname LT7\endcsname{\color{black}}%
      \expandafter\def\csname LT8\endcsname{\color{black}}%
    \fi
  \fi
    \setlength{\unitlength}{0.0500bp}%
    \ifx\gptboxheight\undefined%
      \newlength{\gptboxheight}%
      \newlength{\gptboxwidth}%
      \newsavebox{\gptboxtext}%
    \fi%
    \setlength{\fboxrule}{0.5pt}%
    \setlength{\fboxsep}{1pt}%
\begin{picture}(7200.00,5040.00)%
    \gplgaddtomacro\gplbacktext{%
      \csname LTb\endcsname%
      \put(1078,1076){\makebox(0,0)[r]{\strut{}$1$}}%
      \csname LTb\endcsname%
      \put(1078,2012){\makebox(0,0)[r]{\strut{}$10$}}%
      \csname LTb\endcsname%
      \put(1078,2948){\makebox(0,0)[r]{\strut{}$100$}}%
      \csname LTb\endcsname%
      \put(1078,3883){\makebox(0,0)[r]{\strut{}$1000$}}%
      \csname LTb\endcsname%
      \put(1078,4819){\makebox(0,0)[r]{\strut{}$10000$}}%
      \csname LTb\endcsname%
      \put(1271,484){\makebox(0,0){\strut{}$0.001$}}%
      \csname LTb\endcsname%
      \put(5466,484){\makebox(0,0){\strut{}$0.01$}}%
    }%
    \gplgaddtomacro\gplfronttext{%
      \csname LTb\endcsname%
      \put(209,2761){\rotatebox{-270}{\makebox(0,0){\strut{}$\tau_\text{int}$}}}%
      \put(4006,154){\makebox(0,0){\strut{}$1/N_t$}}%
      \csname LTb\endcsname%
      \put(2926,3256){\makebox(0,0)[r]{\strut{}EFA, $\erwartung{n_\text{el}}$}}%
      \csname LTb\endcsname%
      \put(2926,2926){\makebox(0,0)[r]{\strut{}no FA, $\erwartung{n_\text{el}}$}}%
      \csname LTb\endcsname%
      \put(2926,2596){\makebox(0,0)[r]{\strut{}prediction}}%
      \csname LTb\endcsname%
      \put(2926,2266){\makebox(0,0)[r]{\strut{}no FA, $\erwartung{n_\text{ph}}$}}%
    }%
    \gplbacktext
    \put(0,0){\includegraphics{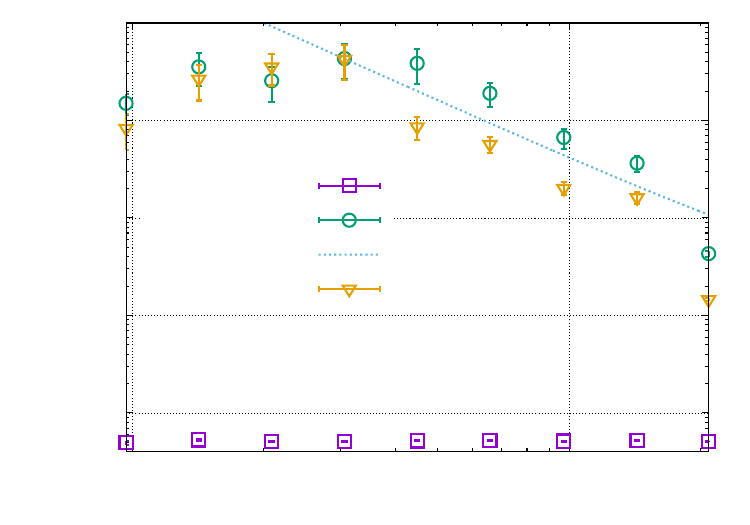}}%
    \gplfronttext
  \end{picture}%
\endgroup
}}
	\caption{Integrated autocorrelation time $\tau_\text{int}$ of the phonon $\erwartung{n_\text{ph}}$ and electron $\erwartung{n_\text{el}}$ number expectation values in the 2+1D SSH model using HMC simulations~\cite{Ostmeyer:2023azi} on a $10\times 10$ lattice with $N_t$ imaginary time slices. With EFA $\tint$ for both observables coincides. The physical parameters are reasonably realistic for the organic semiconductor Rubrene at room temperature $\beta=\SI{40}{\per\eV}$ (see tab.~I of the supl.\ mat.\ in~\cite{Ostmeyer:2023azi} with $\mu=-2\sum_\alpha|J_\alpha|$). Left: different free phonon frequencies $\omega_0$ at constant $N_t=48$ ($\omega_0\approx \SI{6}{\milli\eV}$ is the physical value); Right: different $N_t$ at fixed $\omega_0=\SI{1}{\eV}$. All simulations have similar acceptance ($\gtrsim 80\%$) and compute time per measurement. The dashed line shows the proportionality prediction for $\tau_\text{int}$ from \cref{th:no_fourier_acc}.}
	\label{fig:ssh_rubrene}
\end{figure*}

Figure~\ref{fig:ssh_rubrene} shows the integrated autocorrelation time $\tau_\text{int}$ in HMC simulations of Rubrene, a representative organic molecular semiconductor~\cite{Ostmeyer:2023azi,TroisiNatureMaterials2017,VongJPhysChemLett2022}. Electron numbers are very low in semiconductors $\erwartung{n_\text{el}}\approx1\%$. Thus the electronic interaction term is very small, resulting in mere perturbations from the harmonic phonon action. Therefore, it does not come as a surprise that EFA consistently keeps autocorrelations low $\tau_\text{int}<1$ (completely uncorrelated data has $\tau_\text{int}=\num{0.5}$) throughout all simulations. Without FA we can expect from \cref{th:no_fourier_acc} that $\tint$ will grow as the ratio of largest and smallest eigenvalues of $M_\text{SSH}$
\begin{align}
	\tint &\propto 1 + \left(\frac{2N_t}{\beta\omega_0}\right)^2\,.
\end{align}
Both scenarios of small $\omega_0$ (left) and large $N_t$ (right) are explored in figure~\ref{fig:ssh_rubrene} for two different observables (average electron $\erwartung{n_\text{el}}$ and phonon $\erwartung{n_\text{ph}}$ numbers). These scenarios are highly relevant in practice as the phonon frequency tends to be very small $\omega_0 \ll 1/\beta$ in real materials and the limit $N_t\rightarrow\infty$ is required for the continuum limit. Even though the prefactors in \tint\ clearly depend on the specific observable (as expected), the proportionality from \cref{th:no_fourier_acc} is reproduced in all cases. The plateau in the right panel is an artefact of a limited number of trajectories.

\begin{figure*}[t]
	\centering
	\resizebox{0.98\textwidth}{!}{{\large%
\begingroup
  \inputencoding{latin1}%
  \makeatletter
  \providecommand\color[2][]{%
    \GenericError{(gnuplot) \space\space\space\@spaces}{%
      Package color not loaded in conjunction with
      terminal option `colourtext'%
    }{See the gnuplot documentation for explanation.%
    }{Either use 'blacktext' in gnuplot or load the package
      color.sty in LaTeX.}%
    \renewcommand\color[2][]{}%
  }%
  \providecommand\includegraphics[2][]{%
    \GenericError{(gnuplot) \space\space\space\@spaces}{%
      Package graphicx or graphics not loaded%
    }{See the gnuplot documentation for explanation.%
    }{The gnuplot epslatex terminal needs graphicx.sty or graphics.sty.}%
    \renewcommand\includegraphics[2][]{}%
  }%
  \providecommand\rotatebox[2]{#2}%
  \@ifundefined{ifGPcolor}{%
    \newif\ifGPcolor
    \GPcolortrue
  }{}%
  \@ifundefined{ifGPblacktext}{%
    \newif\ifGPblacktext
    \GPblacktexttrue
  }{}%
  \let\gplgaddtomacro\g@addto@macro
  \gdef\gplbacktext{}%
  \gdef\gplfronttext{}%
  \makeatother
  \ifGPblacktext
    \def\colorrgb#1{}%
    \def\colorgray#1{}%
  \else
    \ifGPcolor
      \def\colorrgb#1{\color[rgb]{#1}}%
      \def\colorgray#1{\color[gray]{#1}}%
      \expandafter\def\csname LTw\endcsname{\color{white}}%
      \expandafter\def\csname LTb\endcsname{\color{black}}%
      \expandafter\def\csname LTa\endcsname{\color{black}}%
      \expandafter\def\csname LT0\endcsname{\color[rgb]{1,0,0}}%
      \expandafter\def\csname LT1\endcsname{\color[rgb]{0,1,0}}%
      \expandafter\def\csname LT2\endcsname{\color[rgb]{0,0,1}}%
      \expandafter\def\csname LT3\endcsname{\color[rgb]{1,0,1}}%
      \expandafter\def\csname LT4\endcsname{\color[rgb]{0,1,1}}%
      \expandafter\def\csname LT5\endcsname{\color[rgb]{1,1,0}}%
      \expandafter\def\csname LT6\endcsname{\color[rgb]{0,0,0}}%
      \expandafter\def\csname LT7\endcsname{\color[rgb]{1,0.3,0}}%
      \expandafter\def\csname LT8\endcsname{\color[rgb]{0.5,0.5,0.5}}%
    \else
      \def\colorrgb#1{\color{black}}%
      \def\colorgray#1{\color[gray]{#1}}%
      \expandafter\def\csname LTw\endcsname{\color{white}}%
      \expandafter\def\csname LTb\endcsname{\color{black}}%
      \expandafter\def\csname LTa\endcsname{\color{black}}%
      \expandafter\def\csname LT0\endcsname{\color{black}}%
      \expandafter\def\csname LT1\endcsname{\color{black}}%
      \expandafter\def\csname LT2\endcsname{\color{black}}%
      \expandafter\def\csname LT3\endcsname{\color{black}}%
      \expandafter\def\csname LT4\endcsname{\color{black}}%
      \expandafter\def\csname LT5\endcsname{\color{black}}%
      \expandafter\def\csname LT6\endcsname{\color{black}}%
      \expandafter\def\csname LT7\endcsname{\color{black}}%
      \expandafter\def\csname LT8\endcsname{\color{black}}%
    \fi
  \fi
    \setlength{\unitlength}{0.0500bp}%
    \ifx\gptboxheight\undefined%
      \newlength{\gptboxheight}%
      \newlength{\gptboxwidth}%
      \newsavebox{\gptboxtext}%
    \fi%
    \setlength{\fboxrule}{0.5pt}%
    \setlength{\fboxsep}{1pt}%
\begin{picture}(7200.00,5040.00)%
    \gplgaddtomacro\gplbacktext{%
      \csname LTb\endcsname%
      \put(946,704){\makebox(0,0)[r]{\strut{}$0.22$}}%
      \csname LTb\endcsname%
      \put(946,1116){\makebox(0,0)[r]{\strut{}$0.24$}}%
      \csname LTb\endcsname%
      \put(946,1527){\makebox(0,0)[r]{\strut{}$0.26$}}%
      \csname LTb\endcsname%
      \put(946,1939){\makebox(0,0)[r]{\strut{}$0.28$}}%
      \csname LTb\endcsname%
      \put(946,2350){\makebox(0,0)[r]{\strut{}$0.3$}}%
      \csname LTb\endcsname%
      \put(946,2762){\makebox(0,0)[r]{\strut{}$0.32$}}%
      \csname LTb\endcsname%
      \put(946,3173){\makebox(0,0)[r]{\strut{}$0.34$}}%
      \csname LTb\endcsname%
      \put(946,3585){\makebox(0,0)[r]{\strut{}$0.36$}}%
      \csname LTb\endcsname%
      \put(946,3996){\makebox(0,0)[r]{\strut{}$0.38$}}%
      \csname LTb\endcsname%
      \put(946,4408){\makebox(0,0)[r]{\strut{}$0.4$}}%
      \csname LTb\endcsname%
      \put(946,4819){\makebox(0,0)[r]{\strut{}$0.42$}}%
      \csname LTb\endcsname%
      \put(1078,484){\makebox(0,0){\strut{}$0$}}%
      \csname LTb\endcsname%
      \put(1794,484){\makebox(0,0){\strut{}$500$}}%
      \csname LTb\endcsname%
      \put(2509,484){\makebox(0,0){\strut{}$1000$}}%
      \csname LTb\endcsname%
      \put(3225,484){\makebox(0,0){\strut{}$1500$}}%
      \csname LTb\endcsname%
      \put(3941,484){\makebox(0,0){\strut{}$2000$}}%
      \csname LTb\endcsname%
      \put(4656,484){\makebox(0,0){\strut{}$2500$}}%
      \csname LTb\endcsname%
      \put(5372,484){\makebox(0,0){\strut{}$3000$}}%
      \csname LTb\endcsname%
      \put(6087,484){\makebox(0,0){\strut{}$3500$}}%
      \csname LTb\endcsname%
      \put(6803,484){\makebox(0,0){\strut{}$4000$}}%
    }%
    \gplgaddtomacro\gplfronttext{%
      \csname LTb\endcsname%
      \put(209,2761){\rotatebox{-270}{\makebox(0,0){\strut{}$\erwartung{n_\text{ph}}$}}}%
      \put(3940,154){\makebox(0,0){\strut{}$t_\text{HMC}$}}%
      \csname LTb\endcsname%
      \put(5816,4591){\makebox(0,0)[r]{\strut{}EFA, $T=\nicefrac{\pi}{2}$}}%
      \csname LTb\endcsname%
      \put(5816,4261){\makebox(0,0)[r]{\strut{}no FA}}%
    }%
    \gplbacktext
    \put(0,0){\includegraphics{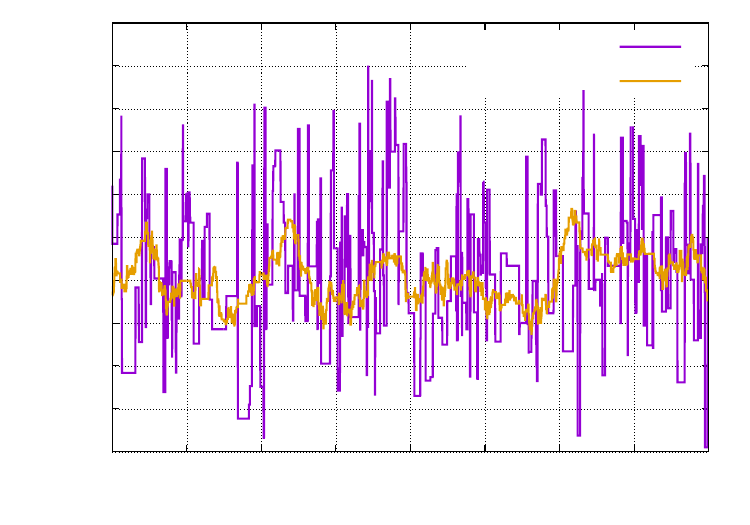}}%
    \gplfronttext
  \end{picture}%
\endgroup
\begingroup
  \inputencoding{latin1}%
  \makeatletter
  \providecommand\color[2][]{%
    \GenericError{(gnuplot) \space\space\space\@spaces}{%
      Package color not loaded in conjunction with
      terminal option `colourtext'%
    }{See the gnuplot documentation for explanation.%
    }{Either use 'blacktext' in gnuplot or load the package
      color.sty in LaTeX.}%
    \renewcommand\color[2][]{}%
  }%
  \providecommand\includegraphics[2][]{%
    \GenericError{(gnuplot) \space\space\space\@spaces}{%
      Package graphicx or graphics not loaded%
    }{See the gnuplot documentation for explanation.%
    }{The gnuplot epslatex terminal needs graphicx.sty or graphics.sty.}%
    \renewcommand\includegraphics[2][]{}%
  }%
  \providecommand\rotatebox[2]{#2}%
  \@ifundefined{ifGPcolor}{%
    \newif\ifGPcolor
    \GPcolortrue
  }{}%
  \@ifundefined{ifGPblacktext}{%
    \newif\ifGPblacktext
    \GPblacktexttrue
  }{}%
  \let\gplgaddtomacro\g@addto@macro
  \gdef\gplbacktext{}%
  \gdef\gplfronttext{}%
  \makeatother
  \ifGPblacktext
    \def\colorrgb#1{}%
    \def\colorgray#1{}%
  \else
    \ifGPcolor
      \def\colorrgb#1{\color[rgb]{#1}}%
      \def\colorgray#1{\color[gray]{#1}}%
      \expandafter\def\csname LTw\endcsname{\color{white}}%
      \expandafter\def\csname LTb\endcsname{\color{black}}%
      \expandafter\def\csname LTa\endcsname{\color{black}}%
      \expandafter\def\csname LT0\endcsname{\color[rgb]{1,0,0}}%
      \expandafter\def\csname LT1\endcsname{\color[rgb]{0,1,0}}%
      \expandafter\def\csname LT2\endcsname{\color[rgb]{0,0,1}}%
      \expandafter\def\csname LT3\endcsname{\color[rgb]{1,0,1}}%
      \expandafter\def\csname LT4\endcsname{\color[rgb]{0,1,1}}%
      \expandafter\def\csname LT5\endcsname{\color[rgb]{1,1,0}}%
      \expandafter\def\csname LT6\endcsname{\color[rgb]{0,0,0}}%
      \expandafter\def\csname LT7\endcsname{\color[rgb]{1,0.3,0}}%
      \expandafter\def\csname LT8\endcsname{\color[rgb]{0.5,0.5,0.5}}%
    \else
      \def\colorrgb#1{\color{black}}%
      \def\colorgray#1{\color[gray]{#1}}%
      \expandafter\def\csname LTw\endcsname{\color{white}}%
      \expandafter\def\csname LTb\endcsname{\color{black}}%
      \expandafter\def\csname LTa\endcsname{\color{black}}%
      \expandafter\def\csname LT0\endcsname{\color{black}}%
      \expandafter\def\csname LT1\endcsname{\color{black}}%
      \expandafter\def\csname LT2\endcsname{\color{black}}%
      \expandafter\def\csname LT3\endcsname{\color{black}}%
      \expandafter\def\csname LT4\endcsname{\color{black}}%
      \expandafter\def\csname LT5\endcsname{\color{black}}%
      \expandafter\def\csname LT6\endcsname{\color{black}}%
      \expandafter\def\csname LT7\endcsname{\color{black}}%
      \expandafter\def\csname LT8\endcsname{\color{black}}%
    \fi
  \fi
    \setlength{\unitlength}{0.0500bp}%
    \ifx\gptboxheight\undefined%
      \newlength{\gptboxheight}%
      \newlength{\gptboxwidth}%
      \newsavebox{\gptboxtext}%
    \fi%
    \setlength{\fboxrule}{0.5pt}%
    \setlength{\fboxsep}{1pt}%
\begin{picture}(7200.00,5040.00)%
    \gplgaddtomacro\gplbacktext{%
      \csname LTb\endcsname%
      \put(946,704){\makebox(0,0)[r]{\strut{}$-0.2$}}%
      \csname LTb\endcsname%
      \put(946,1390){\makebox(0,0)[r]{\strut{}$0$}}%
      \csname LTb\endcsname%
      \put(946,2076){\makebox(0,0)[r]{\strut{}$0.2$}}%
      \csname LTb\endcsname%
      \put(946,2762){\makebox(0,0)[r]{\strut{}$0.4$}}%
      \csname LTb\endcsname%
      \put(946,3447){\makebox(0,0)[r]{\strut{}$0.6$}}%
      \csname LTb\endcsname%
      \put(946,4133){\makebox(0,0)[r]{\strut{}$0.8$}}%
      \csname LTb\endcsname%
      \put(946,4819){\makebox(0,0)[r]{\strut{}$1$}}%
      \csname LTb\endcsname%
      \put(1078,484){\makebox(0,0){\strut{}$1$}}%
      \csname LTb\endcsname%
      \put(2509,484){\makebox(0,0){\strut{}$10$}}%
      \csname LTb\endcsname%
      \put(3941,484){\makebox(0,0){\strut{}$100$}}%
      \csname LTb\endcsname%
      \put(5372,484){\makebox(0,0){\strut{}$1000$}}%
      \csname LTb\endcsname%
      \put(6803,484){\makebox(0,0){\strut{}$10000$}}%
    }%
    \gplgaddtomacro\gplfronttext{%
      \csname LTb\endcsname%
      \put(209,2761){\rotatebox{-270}{\makebox(0,0){\strut{}$\rho_{n_\text{ph}}(\Delta t_\text{HMC})$}}}%
      \put(3940,154){\makebox(0,0){\strut{}$\Delta t_\text{HMC}$}}%
      \csname LTb\endcsname%
      \put(5816,4591){\makebox(0,0)[r]{\strut{}EFA, $T=\nicefrac{\pi}{2}$}}%
      \csname LTb\endcsname%
      \put(5816,4261){\makebox(0,0)[r]{\strut{}EFA, $T=\nicefrac{\pi}{6}$}}%
      \csname LTb\endcsname%
      \put(5816,3931){\makebox(0,0)[r]{\strut{}EFA, $T=\nicefrac{\pi}{20}$}}%
      \csname LTb\endcsname%
      \put(5816,3601){\makebox(0,0)[r]{\strut{}no FA}}%
    }%
    \gplbacktext
    \put(0,0){\includegraphics{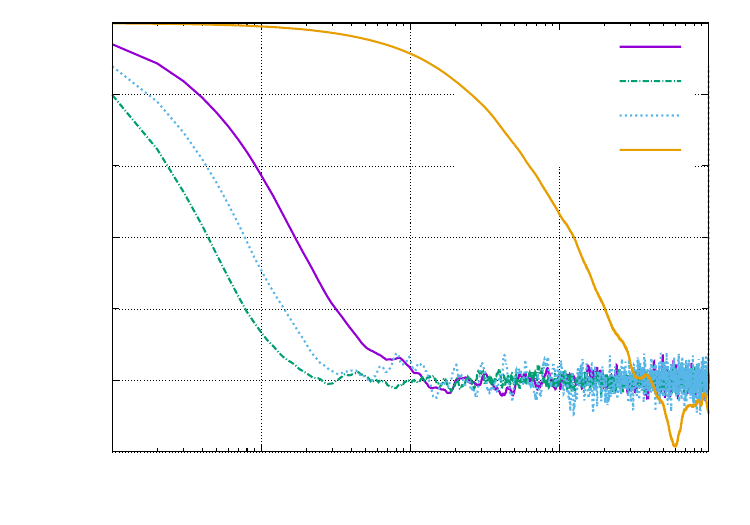}}%
    \gplfronttext
  \end{picture}%
\endgroup
}}
	\caption{Phonon number expectation value $\erwartung{n_\text{ph}}$ (left) and its autocorrelation function (right) in the 2+1D SSH model using HMC simulations~\cite{Ostmeyer:2023azi} on a $10\times 10$ square lattice with $N_t=64$ imaginary time slices. The parameters $\beta=4$, $\omega_0=J_\alpha=g=1$, $\mu=0$ are particularly challenging, such that the system is clearly in the CDW-ordered phase but relatively close to the phase transition~\cite{PhysRevLett.126.017601}. All simulations have similar compute time per measurement, but the acceptance is very low ($<10\%$) for long trajectories $T=\nicefrac\pi2$ because of numerical instabilities. Without FA the integrated autocorrelation time is $\tau_\text{int}=\num{1206\pm327}$, with EFA it is $\tau_\text{int}(\nicefrac{\pi}{20})=\num{8.5\pm.9}$, $\tau_\text{int}(\nicefrac{\pi}{6})=\num{4.5\pm.3}$, $\tau_\text{int}(\nicefrac{\pi}{2})=\num{20.9\pm1.6}$.}
\label{fig:ssh_cdw}
\end{figure*}

The regime investigated for figure~\ref{fig:ssh_cdw} is crucially different from the previous one since these simulations are performed at half filling and the phonon-electron system is strongly interacting. A priori it is not clear how useful FA will be in this regime. However it stands to reason that the slowest phonon modes will still dominate autocorrelation, even though distorted by interactions. Therefore, we should expect EFA to significantly reduce autocorrelation, but not to remove it entirely. This is precisely what can be observed in figure~\ref{fig:ssh_cdw}. It shows the time series of the phonon expectation number and its autocorrelation function for the same physical parameters (in the CDW ordered phase) with and without EFA. Clearly, EFA reduces the autocorrelations by about two orders of magnitude.

The example in figure~\ref{fig:ssh_cdw} is interesting as the optimal trajectory length with EFA appears to be significantly shorter than $T=\nicefrac{\pi}{2}$. The reason in this case is that the simulations involve matrix inverse and determinant calculations of badly conditioned matrices which leads to significant numerical errors.\footnote{The numerical errors can, in principle, be removed by superior algorithms (e.g.\ Schur complement solver~\cite{Ulybyshev_2019}) and/or higher than double floating precision arithmetic. This was deliberately avoided here in order to provide the instructive example.} These errors accumulate more over longer trajectories so that the maximal possible acceptance (even with more MD steps) drops with growing trajectory length. As a consequence, it is beneficial to use shorter trajectories $T\approx\nicefrac\pi6$ in this case because of their higher acceptance and thus lower total autocorrelation. 	
	\subsection{The Ising model}\label{sec:ising}

Now, let us further investigate the importance of choosing the correct trajectory length. For this the Ising model in the HMC formulation of~\cite{ising} is chosen because of its simplicity.

The Ising model describes classical spins $s_i=\pm1$ with the coupling $J$ governed by the Hamiltonian
\begin{align}
	H_\text{Ising} &= -J\sum_{\erwartung{i,j}} s_i s_j\\
	&= -\frac12 J s^\trans K s\,,
\end{align}
where $\erwartung{i,j}$ denotes nearest neighbours. After adding a large enough positive constant $C$ to the connectivity matrix $K$, to make $\tilde K\coloneqq K+C\id$ positive definite, an equivalent action
\begin{align}
	S_\text{Ising} &= -\frac{1}{2J} \phi^\trans \tilde K^{-1} \phi + \sum_i \log\cosh\phi_i
\end{align}
can be derived as in Ref.~\cite{ising}. This action is formulated in terms of continuous fields $\phi$ and it includes a harmonic term, so that \cref{th:opt_traj_and_kin} is applicable for EFA.

In two dimensions the classical Ising model features a phase transition~\cite{onsager_2d_solution}. At low couplings the spins are in a disordered phase and at high coupling they are ferromagnetically ordered. At the phase transition, most Monte Carlo simulation methods experience critical slowing down which means that the integrated autocorrelation time grows as $\tint\sim L^z$ with the lattice size $L$, where the dynamical exponent $z\simeq2$ for the underlying HMC~\cite{ising}. FA cannot overcome critical slowing down because this would require an increased tunnelling probability between different minima of the anharmonic potential while FA acts only on the harmonic components within a local minimum. However, FA can significantly reduce the coefficient in the $L^z$ divergence.

\begin{figure*}[t]
	\centering
	\resizebox{0.98\textwidth}{!}{{\large%
\begingroup
  \inputencoding{latin1}%
  \makeatletter
  \providecommand\color[2][]{%
    \GenericError{(gnuplot) \space\space\space\@spaces}{%
      Package color not loaded in conjunction with
      terminal option `colourtext'%
    }{See the gnuplot documentation for explanation.%
    }{Either use 'blacktext' in gnuplot or load the package
      color.sty in LaTeX.}%
    \renewcommand\color[2][]{}%
  }%
  \providecommand\includegraphics[2][]{%
    \GenericError{(gnuplot) \space\space\space\@spaces}{%
      Package graphicx or graphics not loaded%
    }{See the gnuplot documentation for explanation.%
    }{The gnuplot epslatex terminal needs graphicx.sty or graphics.sty.}%
    \renewcommand\includegraphics[2][]{}%
  }%
  \providecommand\rotatebox[2]{#2}%
  \@ifundefined{ifGPcolor}{%
    \newif\ifGPcolor
    \GPcolortrue
  }{}%
  \@ifundefined{ifGPblacktext}{%
    \newif\ifGPblacktext
    \GPblacktexttrue
  }{}%
  \let\gplgaddtomacro\g@addto@macro
  \gdef\gplbacktext{}%
  \gdef\gplfronttext{}%
  \makeatother
  \ifGPblacktext
    \def\colorrgb#1{}%
    \def\colorgray#1{}%
  \else
    \ifGPcolor
      \def\colorrgb#1{\color[rgb]{#1}}%
      \def\colorgray#1{\color[gray]{#1}}%
      \expandafter\def\csname LTw\endcsname{\color{white}}%
      \expandafter\def\csname LTb\endcsname{\color{black}}%
      \expandafter\def\csname LTa\endcsname{\color{black}}%
      \expandafter\def\csname LT0\endcsname{\color[rgb]{1,0,0}}%
      \expandafter\def\csname LT1\endcsname{\color[rgb]{0,1,0}}%
      \expandafter\def\csname LT2\endcsname{\color[rgb]{0,0,1}}%
      \expandafter\def\csname LT3\endcsname{\color[rgb]{1,0,1}}%
      \expandafter\def\csname LT4\endcsname{\color[rgb]{0,1,1}}%
      \expandafter\def\csname LT5\endcsname{\color[rgb]{1,1,0}}%
      \expandafter\def\csname LT6\endcsname{\color[rgb]{0,0,0}}%
      \expandafter\def\csname LT7\endcsname{\color[rgb]{1,0.3,0}}%
      \expandafter\def\csname LT8\endcsname{\color[rgb]{0.5,0.5,0.5}}%
    \else
      \def\colorrgb#1{\color{black}}%
      \def\colorgray#1{\color[gray]{#1}}%
      \expandafter\def\csname LTw\endcsname{\color{white}}%
      \expandafter\def\csname LTb\endcsname{\color{black}}%
      \expandafter\def\csname LTa\endcsname{\color{black}}%
      \expandafter\def\csname LT0\endcsname{\color{black}}%
      \expandafter\def\csname LT1\endcsname{\color{black}}%
      \expandafter\def\csname LT2\endcsname{\color{black}}%
      \expandafter\def\csname LT3\endcsname{\color{black}}%
      \expandafter\def\csname LT4\endcsname{\color{black}}%
      \expandafter\def\csname LT5\endcsname{\color{black}}%
      \expandafter\def\csname LT6\endcsname{\color{black}}%
      \expandafter\def\csname LT7\endcsname{\color{black}}%
      \expandafter\def\csname LT8\endcsname{\color{black}}%
    \fi
  \fi
    \setlength{\unitlength}{0.0500bp}%
    \ifx\gptboxheight\undefined%
      \newlength{\gptboxheight}%
      \newlength{\gptboxwidth}%
      \newsavebox{\gptboxtext}%
    \fi%
    \setlength{\fboxrule}{0.5pt}%
    \setlength{\fboxsep}{1pt}%
\begin{picture}(7200.00,5040.00)%
    \gplgaddtomacro\gplbacktext{%
      \csname LTb\endcsname%
      \put(814,1117){\makebox(0,0)[r]{\strut{}$1$}}%
      \csname LTb\endcsname%
      \put(814,2489){\makebox(0,0)[r]{\strut{}$10$}}%
      \csname LTb\endcsname%
      \put(814,3860){\makebox(0,0)[r]{\strut{}$100$}}%
      \csname LTb\endcsname%
      \put(946,484){\makebox(0,0){\strut{}$0$}}%
      \csname LTb\endcsname%
      \put(1678,484){\makebox(0,0){\strut{}$0.1$}}%
      \csname LTb\endcsname%
      \put(2410,484){\makebox(0,0){\strut{}$0.2$}}%
      \csname LTb\endcsname%
      \put(3142,484){\makebox(0,0){\strut{}$0.3$}}%
      \csname LTb\endcsname%
      \put(3875,484){\makebox(0,0){\strut{}$0.4$}}%
      \csname LTb\endcsname%
      \put(4607,484){\makebox(0,0){\strut{}$0.5$}}%
      \csname LTb\endcsname%
      \put(5339,484){\makebox(0,0){\strut{}$0.6$}}%
      \csname LTb\endcsname%
      \put(6071,484){\makebox(0,0){\strut{}$0.7$}}%
      \csname LTb\endcsname%
      \put(6803,484){\makebox(0,0){\strut{}$0.8$}}%
    }%
    \gplgaddtomacro\gplfronttext{%
      \csname LTb\endcsname%
      \put(209,2761){\rotatebox{-270}{\makebox(0,0){\strut{}$\tau_\text{int}$}}}%
      \put(3874,154){\makebox(0,0){\strut{}$J$}}%
      \csname LTb\endcsname%
      \put(5816,1592){\makebox(0,0)[r]{\strut{}EFA, $T=\nicefrac\pi6$}}%
      \csname LTb\endcsname%
      \put(5816,1262){\makebox(0,0)[r]{\strut{}EFA, $T=\nicefrac\pi2$}}%
      \csname LTb\endcsname%
      \put(5816,932){\makebox(0,0)[r]{\strut{}prediction for $T=\nicefrac\pi6$}}%
    }%
    \gplbacktext
    \put(0,0){\includegraphics{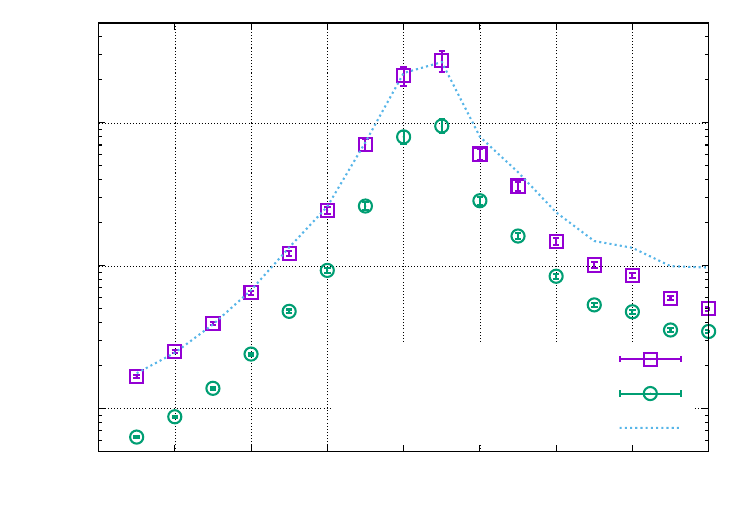}}%
    \gplfronttext
  \end{picture}%
\endgroup
\begingroup
  \inputencoding{latin1}%
  \makeatletter
  \providecommand\color[2][]{%
    \GenericError{(gnuplot) \space\space\space\@spaces}{%
      Package color not loaded in conjunction with
      terminal option `colourtext'%
    }{See the gnuplot documentation for explanation.%
    }{Either use 'blacktext' in gnuplot or load the package
      color.sty in LaTeX.}%
    \renewcommand\color[2][]{}%
  }%
  \providecommand\includegraphics[2][]{%
    \GenericError{(gnuplot) \space\space\space\@spaces}{%
      Package graphicx or graphics not loaded%
    }{See the gnuplot documentation for explanation.%
    }{The gnuplot epslatex terminal needs graphicx.sty or graphics.sty.}%
    \renewcommand\includegraphics[2][]{}%
  }%
  \providecommand\rotatebox[2]{#2}%
  \@ifundefined{ifGPcolor}{%
    \newif\ifGPcolor
    \GPcolortrue
  }{}%
  \@ifundefined{ifGPblacktext}{%
    \newif\ifGPblacktext
    \GPblacktexttrue
  }{}%
  \let\gplgaddtomacro\g@addto@macro
  \gdef\gplbacktext{}%
  \gdef\gplfronttext{}%
  \makeatother
  \ifGPblacktext
    \def\colorrgb#1{}%
    \def\colorgray#1{}%
  \else
    \ifGPcolor
      \def\colorrgb#1{\color[rgb]{#1}}%
      \def\colorgray#1{\color[gray]{#1}}%
      \expandafter\def\csname LTw\endcsname{\color{white}}%
      \expandafter\def\csname LTb\endcsname{\color{black}}%
      \expandafter\def\csname LTa\endcsname{\color{black}}%
      \expandafter\def\csname LT0\endcsname{\color[rgb]{1,0,0}}%
      \expandafter\def\csname LT1\endcsname{\color[rgb]{0,1,0}}%
      \expandafter\def\csname LT2\endcsname{\color[rgb]{0,0,1}}%
      \expandafter\def\csname LT3\endcsname{\color[rgb]{1,0,1}}%
      \expandafter\def\csname LT4\endcsname{\color[rgb]{0,1,1}}%
      \expandafter\def\csname LT5\endcsname{\color[rgb]{1,1,0}}%
      \expandafter\def\csname LT6\endcsname{\color[rgb]{0,0,0}}%
      \expandafter\def\csname LT7\endcsname{\color[rgb]{1,0.3,0}}%
      \expandafter\def\csname LT8\endcsname{\color[rgb]{0.5,0.5,0.5}}%
    \else
      \def\colorrgb#1{\color{black}}%
      \def\colorgray#1{\color[gray]{#1}}%
      \expandafter\def\csname LTw\endcsname{\color{white}}%
      \expandafter\def\csname LTb\endcsname{\color{black}}%
      \expandafter\def\csname LTa\endcsname{\color{black}}%
      \expandafter\def\csname LT0\endcsname{\color{black}}%
      \expandafter\def\csname LT1\endcsname{\color{black}}%
      \expandafter\def\csname LT2\endcsname{\color{black}}%
      \expandafter\def\csname LT3\endcsname{\color{black}}%
      \expandafter\def\csname LT4\endcsname{\color{black}}%
      \expandafter\def\csname LT5\endcsname{\color{black}}%
      \expandafter\def\csname LT6\endcsname{\color{black}}%
      \expandafter\def\csname LT7\endcsname{\color{black}}%
      \expandafter\def\csname LT8\endcsname{\color{black}}%
    \fi
  \fi
    \setlength{\unitlength}{0.0500bp}%
    \ifx\gptboxheight\undefined%
      \newlength{\gptboxheight}%
      \newlength{\gptboxwidth}%
      \newsavebox{\gptboxtext}%
    \fi%
    \setlength{\fboxrule}{0.5pt}%
    \setlength{\fboxsep}{1pt}%
\begin{picture}(7200.00,5040.00)%
    \gplgaddtomacro\gplbacktext{%
      \csname LTb\endcsname%
      \put(682,704){\makebox(0,0)[r]{\strut{}$0$}}%
      \csname LTb\endcsname%
      \put(682,1527){\makebox(0,0)[r]{\strut{}$5$}}%
      \csname LTb\endcsname%
      \put(682,2350){\makebox(0,0)[r]{\strut{}$10$}}%
      \csname LTb\endcsname%
      \put(682,3173){\makebox(0,0)[r]{\strut{}$15$}}%
      \csname LTb\endcsname%
      \put(682,3996){\makebox(0,0)[r]{\strut{}$20$}}%
      \csname LTb\endcsname%
      \put(682,4819){\makebox(0,0)[r]{\strut{}$25$}}%
      \csname LTb\endcsname%
      \put(814,484){\makebox(0,0){\strut{}$0$}}%
      \csname LTb\endcsname%
      \put(1563,484){\makebox(0,0){\strut{}$0.2$}}%
      \csname LTb\endcsname%
      \put(2311,484){\makebox(0,0){\strut{}$0.4$}}%
      \csname LTb\endcsname%
      \put(3060,484){\makebox(0,0){\strut{}$0.6$}}%
      \csname LTb\endcsname%
      \put(3809,484){\makebox(0,0){\strut{}$0.8$}}%
      \csname LTb\endcsname%
      \put(4557,484){\makebox(0,0){\strut{}$1$}}%
      \csname LTb\endcsname%
      \put(5306,484){\makebox(0,0){\strut{}$1.2$}}%
      \csname LTb\endcsname%
      \put(6054,484){\makebox(0,0){\strut{}$1.4$}}%
      \csname LTb\endcsname%
      \put(6803,484){\makebox(0,0){\strut{}$1.6$}}%
    }%
    \gplgaddtomacro\gplfronttext{%
      \csname LTb\endcsname%
      \put(209,2761){\rotatebox{-270}{\makebox(0,0){\strut{}$\tau_\text{int}$}}}%
      \put(3808,154){\makebox(0,0){\strut{}$T$}}%
      \csname LTb\endcsname%
      \put(5816,4591){\makebox(0,0)[r]{\strut{}EFA, $J=\num{0.2}$}}%
      \csname LTb\endcsname%
      \put(5816,4261){\makebox(0,0)[r]{\strut{}prediction}}%
    }%
    \gplbacktext
    \put(0,0){\includegraphics{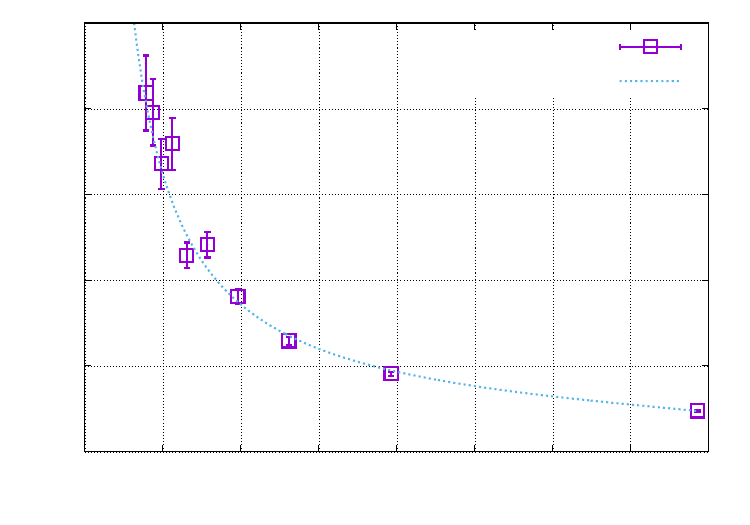}}%
    \gplfronttext
  \end{picture}%
\endgroup
}}
	\caption{Integrated autocorrelation time $\tau_\text{int}$ of the absolute value of the magnetisation $|m|$ in the 2D Ising model using HMC simulations~\cite{ising} on a $15\times 15$ lattice. Left: different coupling strengths ($J\approx \num{0.44}$ is the critical coupling~\cite{onsager_2d_solution}); Right: weak coupling $J=\num{0.2}$ and different trajectory lengths $T$. All simulations used EFA and have similar acceptance $\gtrsim 80\%$. The measurement frequency has been adjusted so that the HMC time between measurements is always the same. The dashed line shows the prediction for $\tau_\text{int}(T)=\tau_\text{int}(T=\nicefrac{\pi}{2})\cdot \left(1-\cos(T)^{\nicefrac{\pi}{2T}}\right)^{-1}$ from \cref{th:tau_int_short_traj}.}
	\label{fig:ising_var_J}
\end{figure*}

The peak of \tint\ around the phase transition at $J\approx \num{0.44}$ is clearly visible in the left panel of figure~\ref{fig:ising_var_J}. It is also clear that the coefficient of \tint\ is larger for a suboptimal trajectory length over the entire parameter range. More specifically, \cref{th:tau_int_short_traj} predicts the coefficient exactly in the weakly coupled phase $J<\num{0.4}$ as expected for the region dominated by the harmonic potential. This is demonstrated even more vividly in the right panel of figure~\ref{fig:ising_var_J}. The quantitative prediction fails in the strongly coupled anharmonic phase $J>\num{0.4}$, but the qualitative advantage of $T=\nicefrac\pi2$ as in \cref{th:opt_traj_and_kin} over $T<\nicefrac\pi2$ is preserved.

In this example it appears that even longer trajectories $T>\nicefrac\pi2$ might reduce \tint\ further. This is a consequence of the significant anharmonic part leading to $\tint \gg \num{0.5}$ even at $T=\nicefrac\pi2$. Since \cref{th:tau_int_short_traj} breaks down for longer trajectories, this phenomenon cannot be predicted analytically. Optimising longer trajectories coupling by coupling would therefore be up to numerical experiments. The advantage, however, is expected to be so small in most cases that it hardly warrants the additional work. 	
	\subsection{Lattice gauge theory}\label{sec:qcd}

Our final example is motivated by lattice quantum chromodynamics (QCD), the theory describing quarks and gluons under the strong interaction. FA has been known in simulations of lattice QCD for decades. The original formulation in Refs.~\cite{DUANE1986143,DUANE1988101,Davies:1989vh} as well as more recent implementations~\cite{Sheta:2021hsd,Huo:2024lns} takes the structure of the underlying gauge group into account explicitly. It cannot be derived from \cref{th:opt_traj_and_kin} and comes at the cost of numerical inversions or solves. In the following, an alternative idea how \cref{th:opt_traj_and_kin} can be used in lattice QCD and gauge theories in general will be sketched. This approximate FA without gauge fixing does not require numerical solves, but it does not mirror the exact group structure either. The approach is a generalisation of the FA used in Ref.~\cite{Borsanyi:2015} for quantum electrodynamics (QED) beyond the abelian U$(1)$ gauge group.

The goal of this section is to convey how broadly applicable \cref{th:opt_traj_and_kin} is, even when it does not appear obvious. This proof-of-concept work does not aim to identify the most efficient simulation method for lattice gauge theories, it merely introduces another alternative to choose from in the future.

Any lattice gauge theory action~\cite{Gattringer:2010zz} is of the form
\begin{align}
	S[U] &= S_G[U] \,+\, \text{fermionic interactions}\,,\\
	S_G[U] &= \frac\beta {N_c}\sum_{n,\mu<\nu} \Re\tr\left[1-U_{\mu\nu}(n)\right]\,,
\end{align}
where the fermionic part will not be considered further. Since the pure gauge part $S_G$ contains the harmonic potential, it will be accelerated in the following. The number of colours $N_c$ in the underlying SU$(N_c)$ or U$(N_c)$ group is left open deliberately. The plaquette
\begin{align}
	U_{\mu\nu}(n) &= U_\mu(n) U_\nu(n+\mu) U_\mu(n+\nu)^\dagger U_\nu(n)^\dagger
\end{align}
is defined in the usual way as a function of the gauge links $U_\mu(n)$. Clearly, the action is not quadratic in the links, so \cref{th:opt_traj_and_kin} is not applicable without further manipulations. Note that the gauge action is quadratic and diagonal in the plaquette $S_G \sim \sum_{n,\mu<\nu} \tr \left[(1-U_{\mu\nu}(n))^\dagger(1-U_{\mu\nu}(n))\right]$ allowing to sample the plaquette directly. This property can be exploited for highly efficient sampling, for instance with the heat bath algorithm~\cite{PhysRevD.21.2308}. The idea there is to update some plaquettes while keeping all but one adjacent links fixed. Then the change of the plaquettes directly translates to a change of the variable link. Unfortunately this ansatz does not generalise to HMC simulations because the EOM for plaquettes and their canonical momenta would inevitably violate unitarity (see \cref{sec:heat_bath} for more details).

In order to avoid such unitarity violations, one has to formulate the action in terms of the algebra rather than group elements, that is
\begin{align}
	U_\mu(n) &= \eto{\im \frac12 x_{n,\mu}\cdot \lambda}\,,\label{eq:link_as_exp}
\end{align}
where $\lambda$ is the vector of generators (Pauli spin matrices $\sigma$ for SU$(2)$, Gell-Mann matrices for SU$(3)$, etc.) and $x$ a vector field. As is derived in \cref{sec:qcd_derivations}, the gauge action can be expanded around small values of $x$ and block-diagonalised via Fourier transformation
\begin{align}
	S_G(x) &= \frac12 x^\trans M_G x + \ordnung{x^4}\,,\\
	&= \frac12 \sum_{k,\mu,\nu} \left[y_{k,\mu}^\dagger\, M^{\mu\nu}_G(k)\, y_{k,\nu}^\pdagger\right] + \ordnung{y^4}\,,\\
	y_{k,\mu} &= \frac{1}{\sqrt{N}}\sum_n \eto{-\im k\cdot n}x_{n,\mu}\,,\\
	M^{\mu\nu}_G(k) &= \frac{2\beta}{N_c} \left[\delta_{\mu\nu}\sum_\rho \sin^2\frac{k_\rho}{2} - \eto{\frac{\im}{2}\left(k_\mu - k_\nu\right)}\sin\frac{k_\mu}{2}\sin\frac{k_\nu}{2}\right]\label{eq:m_of_k_pure_gauge}
\end{align}
with the lattice momenta $k$. The harmonic gauge matrix $M_G$ is defined by its action $M^{\mu\nu}_G(k)$ on the fields $y$ after Fourier transformation. On a hyper-cubic lattice in $d$ dimensions every site has $d$ associated links, so the remaining non-diagonal blocks are of dimensionality $d\times d$ and very fast to diagonalise numerically. At this point \cref{th:opt_traj_and_kin} is directly applicable and momenta can be sampled accordingly. However, the EOM should be integrated numerically by taking the exact links $U_\mu(n)$ into account rather than their harmonic approximations, therefore no EFA is possible. In summary, the FA algorithm employed here only differs from the well-known standard HMC algorithm~\cite{Duane1987,Gattringer:2010zz} in that the kinetic term $\frac12p^\trans \tilde M_G^{-1} p$ is used instead of $\frac12p^2$ (with the regularised generalised inverse $\tilde M_G^{-1}$ defined in \cref{sec:qcd_derivations}). Correctness of the simulations is demonstrated in \cref{sec:lat_gauge_verify}. An alternative approach to FA in gauge theories is sketched in \cref{sec:plaquette_sampling}.

The careful reader might have noticed that there is a minor problem in this formulation because the harmonic gauge matrix $M_G$ is not strictly positive definite. Zero momentum $k=0$ has eigenvalue zero. This makes sense because the gauge action is independent of the global gauge. Furthermore, a complete `band' of zero eigenvalues corresponds to the eigenvectors of the form
\begin{align}
	y^0_{k,\mu} &= \eto{\frac\im2 k_\mu} \sin\frac{k_\mu}{2}.
\end{align}
This band captures the remaining local gauge degrees of freedom. Entirely removing the dynamics of these zero modes, leads to a gauge fixing which is potentially error-prone in non-abelian gauge theories~\cite{DUANE1988101,Huo:2024lns}. The very simple fix used in this work is to replace the exact zeroes by slow dynamics, see \cref{sec:qcd_derivations} for details. An alternative approach could be to regulate the zeros by adding a small constant to $M_G$ the way classical FA typically proceeds. Once all momenta can take non-zero values, the simulation has the same ergodicity properties as the classical HMC without FA and can only differ in terms of efficiency.

\begin{figure*}[t]
	\centering
	\resizebox{0.98\textwidth}{!}{{\large%
\begingroup
  \inputencoding{latin1}%
  \makeatletter
  \providecommand\color[2][]{%
    \GenericError{(gnuplot) \space\space\space\@spaces}{%
      Package color not loaded in conjunction with
      terminal option `colourtext'%
    }{See the gnuplot documentation for explanation.%
    }{Either use 'blacktext' in gnuplot or load the package
      color.sty in LaTeX.}%
    \renewcommand\color[2][]{}%
  }%
  \providecommand\includegraphics[2][]{%
    \GenericError{(gnuplot) \space\space\space\@spaces}{%
      Package graphicx or graphics not loaded%
    }{See the gnuplot documentation for explanation.%
    }{The gnuplot epslatex terminal needs graphicx.sty or graphics.sty.}%
    \renewcommand\includegraphics[2][]{}%
  }%
  \providecommand\rotatebox[2]{#2}%
  \@ifundefined{ifGPcolor}{%
    \newif\ifGPcolor
    \GPcolortrue
  }{}%
  \@ifundefined{ifGPblacktext}{%
    \newif\ifGPblacktext
    \GPblacktexttrue
  }{}%
  \let\gplgaddtomacro\g@addto@macro
  \gdef\gplbacktext{}%
  \gdef\gplfronttext{}%
  \makeatother
  \ifGPblacktext
    \def\colorrgb#1{}%
    \def\colorgray#1{}%
  \else
    \ifGPcolor
      \def\colorrgb#1{\color[rgb]{#1}}%
      \def\colorgray#1{\color[gray]{#1}}%
      \expandafter\def\csname LTw\endcsname{\color{white}}%
      \expandafter\def\csname LTb\endcsname{\color{black}}%
      \expandafter\def\csname LTa\endcsname{\color{black}}%
      \expandafter\def\csname LT0\endcsname{\color[rgb]{1,0,0}}%
      \expandafter\def\csname LT1\endcsname{\color[rgb]{0,1,0}}%
      \expandafter\def\csname LT2\endcsname{\color[rgb]{0,0,1}}%
      \expandafter\def\csname LT3\endcsname{\color[rgb]{1,0,1}}%
      \expandafter\def\csname LT4\endcsname{\color[rgb]{0,1,1}}%
      \expandafter\def\csname LT5\endcsname{\color[rgb]{1,1,0}}%
      \expandafter\def\csname LT6\endcsname{\color[rgb]{0,0,0}}%
      \expandafter\def\csname LT7\endcsname{\color[rgb]{1,0.3,0}}%
      \expandafter\def\csname LT8\endcsname{\color[rgb]{0.5,0.5,0.5}}%
    \else
      \def\colorrgb#1{\color{black}}%
      \def\colorgray#1{\color[gray]{#1}}%
      \expandafter\def\csname LTw\endcsname{\color{white}}%
      \expandafter\def\csname LTb\endcsname{\color{black}}%
      \expandafter\def\csname LTa\endcsname{\color{black}}%
      \expandafter\def\csname LT0\endcsname{\color{black}}%
      \expandafter\def\csname LT1\endcsname{\color{black}}%
      \expandafter\def\csname LT2\endcsname{\color{black}}%
      \expandafter\def\csname LT3\endcsname{\color{black}}%
      \expandafter\def\csname LT4\endcsname{\color{black}}%
      \expandafter\def\csname LT5\endcsname{\color{black}}%
      \expandafter\def\csname LT6\endcsname{\color{black}}%
      \expandafter\def\csname LT7\endcsname{\color{black}}%
      \expandafter\def\csname LT8\endcsname{\color{black}}%
    \fi
  \fi
    \setlength{\unitlength}{0.0500bp}%
    \ifx\gptboxheight\undefined%
      \newlength{\gptboxheight}%
      \newlength{\gptboxwidth}%
      \newsavebox{\gptboxtext}%
    \fi%
    \setlength{\fboxrule}{0.5pt}%
    \setlength{\fboxsep}{1pt}%
    \definecolor{tbcol}{rgb}{1,1,1}%
\begin{picture}(7200.00,5040.00)%
    \gplgaddtomacro\gplbacktext{%
      \csname LTb\endcsname%
      \put(814,1242){\makebox(0,0)[r]{\strut{}$1$}}%
      \csname LTb\endcsname%
      \put(814,3031){\makebox(0,0)[r]{\strut{}$10$}}%
      \csname LTb\endcsname%
      \put(814,4819){\makebox(0,0)[r]{\strut{}$100$}}%
      \csname LTb\endcsname%
      \put(1191,484){\makebox(0,0){\strut{}$20$}}%
      \csname LTb\endcsname%
      \put(1805,484){\makebox(0,0){\strut{}$40$}}%
      \csname LTb\endcsname%
      \put(2418,484){\makebox(0,0){\strut{}$60$}}%
      \csname LTb\endcsname%
      \put(3031,484){\makebox(0,0){\strut{}$80$}}%
      \csname LTb\endcsname%
      \put(3645,484){\makebox(0,0){\strut{}$100$}}%
      \csname LTb\endcsname%
      \put(4258,484){\makebox(0,0){\strut{}$120$}}%
      \csname LTb\endcsname%
      \put(4871,484){\makebox(0,0){\strut{}$140$}}%
      \csname LTb\endcsname%
      \put(5484,484){\makebox(0,0){\strut{}$160$}}%
      \csname LTb\endcsname%
      \put(6098,484){\makebox(0,0){\strut{}$180$}}%
      \csname LTb\endcsname%
      \put(6711,484){\makebox(0,0){\strut{}$200$}}%
    }%
    \gplgaddtomacro\gplfronttext{%
      \csname LTb\endcsname%
      \put(4305,4591){\makebox(0,0)[r]{\strut{}FA, $T=\nicefrac\pi2$}}%
      \csname LTb\endcsname%
      \put(4305,4261){\makebox(0,0)[r]{\strut{}no FA, $T=\num{0.6}$}}%
      \csname LTb\endcsname%
      \put(209,2761){\rotatebox{-270.00}{\makebox(0,0){\strut{}$\tau_\text{int}$}}}%
      \put(3874,154){\makebox(0,0){\strut{}$L$}}%
    }%
    \gplbacktext
    \put(0,0){\includegraphics[width={360.00bp},height={252.00bp}]{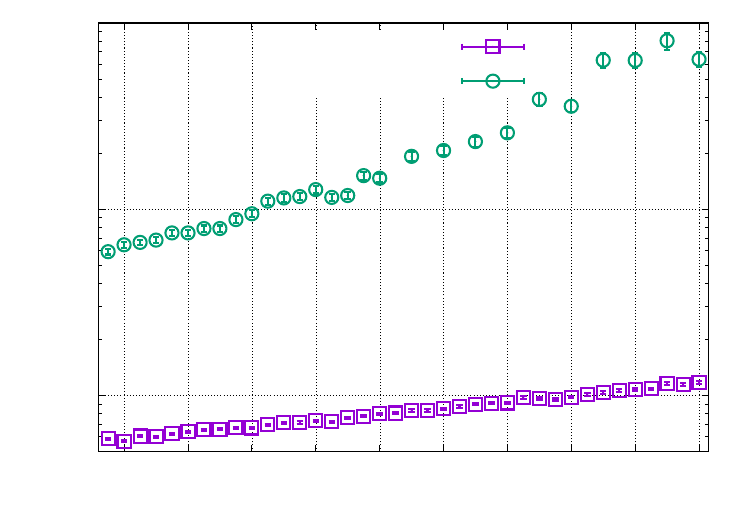}}%
    \gplfronttext
  \end{picture}%
\endgroup
\begingroup
  \inputencoding{latin1}%
  \makeatletter
  \providecommand\color[2][]{%
    \GenericError{(gnuplot) \space\space\space\@spaces}{%
      Package color not loaded in conjunction with
      terminal option `colourtext'%
    }{See the gnuplot documentation for explanation.%
    }{Either use 'blacktext' in gnuplot or load the package
      color.sty in LaTeX.}%
    \renewcommand\color[2][]{}%
  }%
  \providecommand\includegraphics[2][]{%
    \GenericError{(gnuplot) \space\space\space\@spaces}{%
      Package graphicx or graphics not loaded%
    }{See the gnuplot documentation for explanation.%
    }{The gnuplot epslatex terminal needs graphicx.sty or graphics.sty.}%
    \renewcommand\includegraphics[2][]{}%
  }%
  \providecommand\rotatebox[2]{#2}%
  \@ifundefined{ifGPcolor}{%
    \newif\ifGPcolor
    \GPcolortrue
  }{}%
  \@ifundefined{ifGPblacktext}{%
    \newif\ifGPblacktext
    \GPblacktexttrue
  }{}%
  \let\gplgaddtomacro\g@addto@macro
  \gdef\gplbacktext{}%
  \gdef\gplfronttext{}%
  \makeatother
  \ifGPblacktext
    \def\colorrgb#1{}%
    \def\colorgray#1{}%
  \else
    \ifGPcolor
      \def\colorrgb#1{\color[rgb]{#1}}%
      \def\colorgray#1{\color[gray]{#1}}%
      \expandafter\def\csname LTw\endcsname{\color{white}}%
      \expandafter\def\csname LTb\endcsname{\color{black}}%
      \expandafter\def\csname LTa\endcsname{\color{black}}%
      \expandafter\def\csname LT0\endcsname{\color[rgb]{1,0,0}}%
      \expandafter\def\csname LT1\endcsname{\color[rgb]{0,1,0}}%
      \expandafter\def\csname LT2\endcsname{\color[rgb]{0,0,1}}%
      \expandafter\def\csname LT3\endcsname{\color[rgb]{1,0,1}}%
      \expandafter\def\csname LT4\endcsname{\color[rgb]{0,1,1}}%
      \expandafter\def\csname LT5\endcsname{\color[rgb]{1,1,0}}%
      \expandafter\def\csname LT6\endcsname{\color[rgb]{0,0,0}}%
      \expandafter\def\csname LT7\endcsname{\color[rgb]{1,0.3,0}}%
      \expandafter\def\csname LT8\endcsname{\color[rgb]{0.5,0.5,0.5}}%
    \else
      \def\colorrgb#1{\color{black}}%
      \def\colorgray#1{\color[gray]{#1}}%
      \expandafter\def\csname LTw\endcsname{\color{white}}%
      \expandafter\def\csname LTb\endcsname{\color{black}}%
      \expandafter\def\csname LTa\endcsname{\color{black}}%
      \expandafter\def\csname LT0\endcsname{\color{black}}%
      \expandafter\def\csname LT1\endcsname{\color{black}}%
      \expandafter\def\csname LT2\endcsname{\color{black}}%
      \expandafter\def\csname LT3\endcsname{\color{black}}%
      \expandafter\def\csname LT4\endcsname{\color{black}}%
      \expandafter\def\csname LT5\endcsname{\color{black}}%
      \expandafter\def\csname LT6\endcsname{\color{black}}%
      \expandafter\def\csname LT7\endcsname{\color{black}}%
      \expandafter\def\csname LT8\endcsname{\color{black}}%
    \fi
  \fi
    \setlength{\unitlength}{0.0500bp}%
    \ifx\gptboxheight\undefined%
      \newlength{\gptboxheight}%
      \newlength{\gptboxwidth}%
      \newsavebox{\gptboxtext}%
    \fi%
    \setlength{\fboxrule}{0.5pt}%
    \setlength{\fboxsep}{1pt}%
    \definecolor{tbcol}{rgb}{1,1,1}%
\begin{picture}(7200.00,5040.00)%
    \gplgaddtomacro\gplbacktext{%
      \csname LTb\endcsname%
      \put(682,1477){\makebox(0,0)[r]{\strut{}$1$}}%
      \csname LTb\endcsname%
      \put(682,4046){\makebox(0,0)[r]{\strut{}$10$}}%
      \csname LTb\endcsname%
      \put(2421,484){\makebox(0,0){\strut{}$3$}}%
      \csname LTb\endcsname%
      \put(3435,484){\makebox(0,0){\strut{}$6$}}%
      \csname LTb\endcsname%
      \put(5789,484){\makebox(0,0){\strut{}$30$}}%
      \csname LTb\endcsname%
      \put(814,484){\makebox(0,0){\strut{}$1$}}%
      \csname LTb\endcsname%
      \put(4182,484){\makebox(0,0){\strut{}$10$}}%
    }%
    \gplgaddtomacro\gplfronttext{%
      \csname LTb\endcsname%
      \put(2926,4591){\makebox(0,0)[r]{\strut{}FA, $T=\nicefrac\pi2$}}%
      \csname LTb\endcsname%
      \put(2926,4261){\makebox(0,0)[r]{\strut{}no FA, $T=\nicefrac{1}{\sqrt{\beta}}$  }}%
      \csname LTb\endcsname%
      \put(2926,3931){\makebox(0,0)[r]{\strut{}no FA, $T=1$}}%
      \csname LTb\endcsname%
      \put(209,2761){\rotatebox{-270.00}{\makebox(0,0){\strut{}$\tau_\text{int}$}}}%
      \put(3808,154){\makebox(0,0){\strut{}$\beta$}}%
    }%
    \gplbacktext
    \put(0,0){\includegraphics[width={360.00bp},height={252.00bp}]{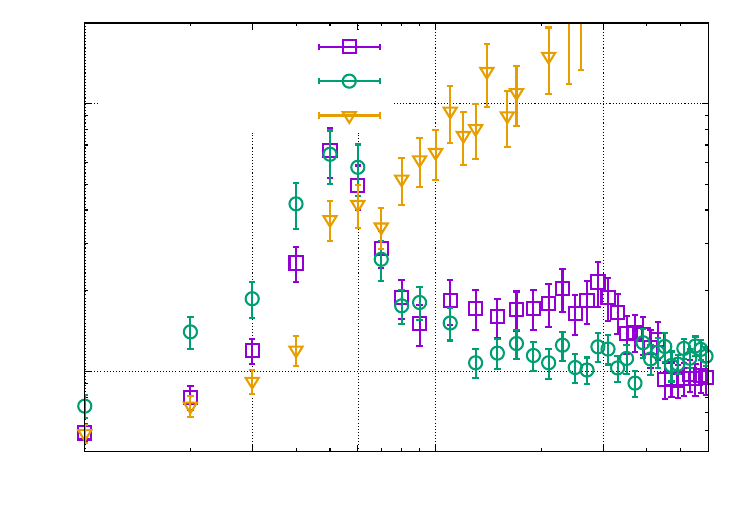}}%
    \gplfronttext
  \end{picture}%
\endgroup
}}
	\caption{Integrated autocorrelation time $\tau_\text{int}$ of the plaquette expectation value $\erwartung{P(\beta)}$ in pure gauge theory HMC simulations. Left: 2D, U$(1)$ weak coupling $\beta=\num{10}$ and different lattice sizes $L$; Right: 4D, $L=10$ lattice, SU$(3)$ and different coupling strengths $\beta$. All simulations required the same compute time per trajectory and volume. Measurements were performed every trajectory. Trajectory lengths without FA were chosen as follows. Left: $T=\num{0.6}$ tuned (by hand) to minimise $\tau_\text{int}$ on the $15\times15$ lattice; Right: $T=1$ because it is the `canonical' choice and $T=\nicefrac{1}{\sqrt{\beta}}$ because of its correct scaling from \cref{th:opt_traj_and_kin}.}
	\label{fig:pure_gauge_tau}
\end{figure*}

As derived in \cref{sec:qcd_derivations}, the approximate FA is guaranteed to negate autocorrelation in the weak coupling (large $\beta$) regime. Both panels of figure~\ref{fig:pure_gauge_tau} clearly show this behaviour. In particular, in the left panel 2D simulations of the U$(1)$ gauge group at weak coupling are presented for different lattice sizes. This is the regime required for realistic QED simulations and FA has proved very effective in this special case~\cite{Borsanyi:2015}. FA has an integrated autocorrelation time of the plaquette well below $\tint<2$ for all simulated sizes which is much lower than in all simulations without FA. In addition, \tint\ with FA depends only very mildly on the system size compared to \tint\ without FA (i.e.\ the slope on the log-scale is lower). This dependence is primarily caused by a decreasing acceptance as $L$ increases and it could be alleviated by either shortening the trajectory length (also increasing autocorrelation) or scaling the number of MD steps accordingly (increasing computation time per trajectory). Neither adjustment is necessary with FA. A further advantage of FA is that it does not require any tuning of the trajectory length while in the simulations without FA $T=\num{0.6}$ had to be tuned by hand in order to avoid even higher autocorrelation.

Just as in the Ising model simulations described in \cref{sec:ising}, FA cannot remove autocorrelation intrinsically caused by anharmonic properties of the potential. For lattice gauge theories the so-called topological freezing can pose a severe problem~\cite{Schaefer:2010hu}. Again, FA does not remove this effect, it only reduces the overall coefficient.

The right panel of figure~\ref{fig:pure_gauge_tau} shows the integrated autocorrelation time in QCD-motivated simulations, i.e.\ in 4D with the SU$(3)$ gauge group. Simulations were performed on a relatively small lattice of length $L=10$ as a function of the coupling $\beta$. As expected, simulations with FA approach minimal autocorrelation for large $\beta>30$. \tint\ with FA has a maximum at intermediate coupling $\beta\approx5$. These results are compared to simulations without FA using fixed $T=1$ and adjusted $T=\nicefrac{1}{\sqrt{\beta}}$. The latter is already using insight from \cref{th:opt_traj_and_kin}, so technically it is harmonically accelerated, even though it does not rely on FA in the narrow sense.

The strong coupling limit $\beta\lesssim 1$ is very easy to simulate with and without FA, always yielding low autocorrelation. All simulations show an increase of \tint\ towards intermediate coupling. The fixed trajectory length $T=1$ appears preferable below $\beta\lesssim6$. However, this is an artefact of the excessively high number of molecular dynamics steps chosen here in order to observe the true decorrelation decoupled from acceptance influence. As long as the acceptance with $T=1$ is high enough, the longer trajectory allows better decorrelation. For weaker couplings $\beta>6$ (earlier in simulations with a realistic number of MD steps) this advantage vanishes because the acceptance goes to zero. In order to maintain a high acceptance at fixed $T=1$, one would need to scale the number of MD steps and thus computational effort with $\sqrt{\beta}$ which is clearly less efficient than to reduce the trajectory length by the same factor.

Qualitatively, \tint\ behaves very similarly in simulations with FA and those without FA but with adjusted $T=\nicefrac{1}{\sqrt{\beta}}$. A priori this is not clear and it might well be that this similarity does not extend to observables other than the plaquette. The scaling with system size should favour FA, similarly to the simpler 2D example in the left panel of figure~\ref{fig:pure_gauge_tau}. For the given observable and lattice size there are still several small but significant differences. This approximate FA has been designed to work best in the weak coupling limit, accordingly it outperforms simulations without FA for large $\beta>50$. On the other hand, weak but not extreme coupling $10\lesssim \beta\lesssim40$ favours simulations without FA. In this region, simulations with FA also took very many trajectories $\ordnung{1000}$ to thermalise. The origin of this bad performance is not yet understood. On a side note, decorrelation per accepted trajectory is usually slightly better with FA than without, but this is compensated by a slightly higher acceptance of simulations without FA and adjusted $T=\nicefrac{1}{\sqrt{\beta}}$.

Finally, \cref{th:opt_traj_and_kin} only demands the proportionality $T \sqrt{\beta}=\text{const.}$, but the optimal coefficient cannot be fixed without further tuning (unless full FA is used). It is therefore very likely that $T=\nicefrac{1}{\sqrt{\beta}}$ is not even close to optimal. Moreover, the autocorrelation is strongly observable dependent, so that the best choice of $T$ might also depend on the observable of interest.
For instance, in Ref.~\cite{MEYER200791} the optimal trajectory length for slowly decorrelating observables at $\beta\approx6$ is found empirically to be around $T\approx2$, implying that $T=\nicefrac{5}{\sqrt{\beta}}$ might be a good choice.

Overall, the benchmark results are not conclusive and more testing, including comparisons to alternative simulation methods, will be needed. The optimal algorithm choice clearly depends the physical parameters. For abelian groups and for very weak couplings as in QED the approximate FA appears viable. The physically relevant coupling in QCD lies around $\beta\approx 6$, just in the region of maximal \tint. Thus, in this and similar regimes methods like the classical FA~\cite{DUANE1986143,DUANE1988101} or no FA at all might be preferable. 	
	\section{Conclusion}\label{sec:conclusion}

In this work, the optimal parameters for hybrid Monte Carlo (HMC) simulations are explored. \Cref{th:opt_traj_and_kin} presents a way to completely eliminate autocorrelations stemming from harmonic contributions of the action. The method relies on the analytic solution of the harmonic oscillator equations and is called exact Fourier acceleration (EFA), consistently with our previous (sub-optimal) formulation in Ref.~\cite{Ostmeyer:2023azi}. It is summarised in  \cref{alg:efa,alg:efa-hmc,alg:leap-frog}.

Only the combination of all the following points, collected in the EFA approach for the first time, allows for optimal sampling:
\begin{itemize}
	\item kinetic `mass' matrix $M^{-1}$ (not only for Fourier invertible cases),
	\item HMC trajectory length $T=\frac\pi2$,
	\item exact integration of the harmonic forces, numerical treatment of the rest.
\end{itemize}
With EFA the HMC becomes equivalent to direct sampling for harmonic actions $S(x)=\frac12 x^\trans M x$ and only perturbatively worse for perturbative anharmonic parts.

Since usually little to no analytic insight is available for anharmonic contributions to the action, EFA remains the best available method for most physical systems, far beyond the harmonically dominated case. One way to intuitively understand it, is that HMC with EFA reconciles the classical HMC ansatz with optimal sampling of normally distributed random variables. That is, for a purely harmonic action the HMC following \cref{th:opt_traj_and_kin} coincides with uncorrelated sampling from a normal distribution, naturally generalising to more complicated actions without sacrificing performance.

The use of EFA often involves an FFT and always an adjustment of the HMC trajectory length. Both measures are computationally very cheap. Neglecting either of these adjustments can lead to an increase of the integrated autocorrelation time \tint\ in simulations of physically relevant systems by many orders of magnitude (see figs.~\ref{fig:ssh_rubrene}, \ref{fig:ising_var_J}). The functional dependences of \tint\ on the condition number of the harmonic matrix and the trajectory length have been quantified in \cref{th:no_fourier_acc,th:tau_int_short_traj}, respectively. Both theoretical predictions have been found in excellent agreement with data obtained in numerical simulations of the SSH and the Ising models.

In \cref{sec:qcd} an approximate form of Fourier acceleration (FA) has been applied to lattice gauge theory. It worked very well in the abelian case, relevant for quantum electrodynamics (QED), but the results for non-abelian gauge groups, required e.g.\ for quantum chromodynamics (QCD), remain inconclusive. Apart from the very weakly coupled regime where the FA becomes exact, it is not clear whether FA can outperform more conventional methods. The answer appears parameter-dependent and will require further case-by-case tuning or, ideally, an improved version of the FA. Nevertheless, the insight from \cref{th:opt_traj_and_kin} still proved useful for the simulation of lattice gauge theories, as it provides us with the optimal scaling of the trajectory length with the coupling $T\propto \nicefrac{1}{\sqrt{\beta}}$.

For all its merits, EFA does not help the HMC to avoid critical slowing down because it only optimises sampling within a local minimum of the action. EFA does not introduce any mechanism to overcome potential barriers (see fig.~\ref{fig:ising_var_J}, left). In this sense it is neither advantageous nor disadvantageous compared to classical HMC. For systems with de-facto ergodicity problems like topological freezing additional methods have to be combined with EFA in order to guarantee efficient sampling. Finding these methods is, of course, a very challenging task and the search will go on. 	
	\section*{Code and Data}
	The simulations of the Ising model in \cref{sec:ising} have been performed with a modified version of the code~\cite{ising_HMC} published with Ref.~\cite{ising}. The code for the SSH model simulations in \cref{sec:ssh} has been published recently~\cite{ssh_simulations}, accompanying an updated version of Ref.~\cite{Ostmeyer:2023azi}. The lattice gauge theory code~\cite{gauge_simulations} used in \cref{sec:qcd} has been written from scratch for this work. The analysis used the light-weight tool \texttt{comp-avg}~\cite{comp-avg}. All these codes are implemented in \texttt{C} and have been published under open access. Most of the simulations in this work can be reproduced very quickly, nonetheless the resulting data will gladly be provided upon request.
	
	\section*{Acknowledgements}
	We thank Evan Berkowitz, Benjamin Cohen-Stead, Sander Gribling, Anthony Kennedy, Stefan Krieg, Tom Luu, and Carsten Urbach for their helpful comments.
	Marcel Rodekamp deserves special thanks for the very enjoyable and productive joint literature research of pure gauge theory and topology.
	This work was funded in part by the STFC Consolidated Grant ST/T000988/1 and by the Deutsche Forschungsgemeinschaft (DFG, German Research Foundation) as part of the CRC 1639/1 NuMeriQS – 511713970.
	Numerical simulations were undertaken on Barkla (though simulations with EFA could have easily been run on a laptop), part of the High Performance Computing facilities at the University of Liverpool, UK.

	\printbibliography

	\allowdisplaybreaks[1]
	\appendix

\section[Proofs of the corollaries]{Proofs of \cref{th:no_fourier_acc,th:tau_int_short_traj}}\label{sec:proofs}

\begin{proof}[Proof: no Fourier acceleration, \cref{th:no_fourier_acc}]
	In order to maintain a constant acceptance rate and computational cost per trajectory, the trajectory length has to be chosen so that $\omega_\mathrm{max} T=\text{const.}$ because (unless in a null set of cases some periodic behaviour is achieved) the fastest mode is responsible for the biggest changes after a fixed (short) time and therefore dominates the acceptance. We set
	\begin{align}
		T &= \frac{\alpha}{\omega_\mathrm{max}}
	\end{align}
	with some positive constant $\alpha$. Autocorrelation, on the other hand, is primarily caused by the slowest mode that requires most HMC steps to change significantly. The autocorrelation of the configuration $x$ is thus dominated by
	\begin{align}
		\frac{\erwartung{x_{i_\text{min}}\left(\frac{\alpha}{\omega_\mathrm{max}}\right)\,x_{i_\text{min}}^0}}{\erwartung{\left(x_{i_\text{min}}^0\right)^2}} &= \frac{\erwartung{x_{i_\text{min}}^0\,x_{i_\text{min}}^0}}{\erwartung{\left(x_{i_\text{min}}^0\right)^2}} \cos\left(\alpha\frac{\omega_\text{min}}{\omega_\mathrm{max}}\right) + \frac{\erwartung{p_{i_\text{min}}^0\,x_{i_\text{min}}^0}}{\erwartung{\left(x_{i_\text{min}}^0\right)^2}} \frac{\sin\left(\alpha\frac{\omega_\text{min}}{\omega_\mathrm{max}}\right)}{\omega_\text{min}}\\
		&= \cos\left(\alpha\frac{\omega_\text{min}}{\omega_\mathrm{max}}\right)\,,\label{eq:corr_traj_omega}
	\end{align}
	where $i_\text{min}$ is the index corresponding to the slowest mode $\omega_\mathrm{min}$. For $x(T)$ the explicit formula~\eqref{eq:eom_sol_x_of_t} has been used. The sin-term can be dropped because $x^0$ and $p^0$ are uncorrelated, i.e.\ $\erwartung{x_i^0 p_i^0} = \erwartung{x_i^0}\erwartung{p_i^0}$, and $\erwartung{p^0}=0$.
	
	The correlation after $n$ trajectories is given by the power $\cos\left(\alpha\frac{\omega_\text{min}}{\omega_\mathrm{max}}\right)^n$ and the integrated autocorrelation time is directly proportional to the geometric series over the correlation function
	\begin{align}
		\tau_\mathrm{int} &\propto \sum_{n=0}^{\infty}\cos\left(\alpha\frac{\omega_\text{min}}{\omega_\mathrm{max}}\right)^n \\
		&= \left(1-\cos\left(\alpha\frac{\omega_\mathrm{min}}{\omega_\mathrm{max}}\right)\right)^{-1} \\
		&= 2\left(\alpha^{-1}\frac{\omega_\mathrm{max}}{\omega_\mathrm{min}}\right)^2 + \frac16 + \ordnung{\left(\frac{\omega_\mathrm{min}}{\omega_\mathrm{max}}\right)^2}\,.
	\end{align}
\end{proof}

\begin{proof}[Proof: too short trajectory, \cref{th:tau_int_short_traj}]
	The proof proceeds very similarly to that of \cref{th:no_fourier_acc} with two differences. First of all the FA rescales all frequencies $\omega_i$ by their inverse, so the correlation~\eqref{eq:corr_traj_omega} after a single trajectory simplifies to
	\begin{align}
		\frac{\erwartung{x_{i}\left(T\right)\,x_{i}^0}}{\erwartung{\left(x_{i}^0\right)^2}} &= \cos(T)
	\end{align}
	for all $i$. Secondly, a shorter trajectory comes with reduced computational cost, so a fair comparison is only ensured when proportionally rarer measurements are conducted. The fictitious HMC time between measurements (equivalent to the computational cost) is kept fixed by measuring only every $\frac{\pi}{2T}$-th trajectory with respect to the measurement frequency of $T=\frac\pi2$. After $\frac{\pi}{2T}$ trajectories the correlation is given by $\cos(T)^{\nicefrac{\pi}{2T}}$ and the geometric series therefore amounts to
	\begin{align}
		\tau_\mathrm{int}(T) &= \tau_\mathrm{int}(T=\nicefrac{\pi}{2}) \sum_{n=0}^\infty \cos(T)^{n\nicefrac{\pi}{2T}}\\
		&= \frac{\tau_\mathrm{int}(T=\nicefrac{\pi}{2})}{ 1-\cos(T)^{\nicefrac{\pi}{2T}}}\,.
	\end{align}
\end{proof} 	
	\section{The Hubbard model}\label{sec:hubbard_model}

The Hubbard model~\cite{Hubbard1963} is mentioned here as an example of pure ``harmonic'' acceleration without the need to perform a Fourier transformation since the matrix $M\propto\id$ in the action~\eqref{eq:general_action} is already diagonal in real space.

The Hamiltonian defining the Hubbard model reads
\begin{align}
	H_\text{Hub} &= -\kappa \sum_{\erwartung{x,y},\sigma} \left(c^\dagger_{x,\sigma}c^\pdagger_{y,\sigma} + c^\dagger_{y,\sigma}c^\pdagger_{x,\sigma}\right) + U \sum_x n_{x\uparrow}n_{x\downarrow}\,,
\end{align}
where $\kappa$ denotes the hopping amplitude between nearest neighbours $\erwartung{x,y}$ and $U$ is the (repulsive) on-site interaction strength. The creation (annihilation) operators $c^\dagger_{x,\sigma}$ ($c^\pdagger_{x,\sigma}$) and corresponding number operators $n_{x,\sigma}\equiv c^\dagger_{x,\sigma}c^\pdagger_{x,\sigma}$ represent fermions (e.g.\ electrons) of spin $\sigma$ at site $x$.

Following the formalism of Refs.~\cite{PhysRevB.93.155106,acceleratingHMC,semimetalmott} and setting $\tilde U\coloneqq\frac{\beta U}{N_t}$, the grand-canonical action
\begin{align}
	S_\text{Hub}(\phi) &= \frac{1}{2\tilde U}\phi^2 + \text{interactions}\\ %
	\xRightarrow{\text{th.~\ref{th:opt_traj_and_kin}}} \mathcal{H}_\text{Hub} &= \frac{\tilde U}{2} p^2 + S_\text{Hub}(\phi)
\end{align}
at inverse temperature $\beta$ can be derived. The number of Euclidean time slices $N_t$ has to be large $N_t\gg \beta U$ in order to approach the continuum limit, so typically the `mass' of the canonical momenta $p$ associated with the Hubbard–Stratonovich fields $\phi$ should be chosen larger than unity as well.

As mentioned before, the harmonic matrix $M_\text{Hub} = \frac{1}{2\tilde U}\id$ is proportional to the identity. Therefore, all masses should be chosen equal according to \cref{th:opt_traj_and_kin}. In fact, for simplicity's sake, the momenta $p$ can be sampled from a standard normal distribution $\mathcal{H}_\text{Hub} = \frac12p^2 + S_\text{Hub}(\phi)$ with the only degree of freedom absorbed into the trajectory length $T=\frac\pi2 \sqrt{\tilde U}$. 	
\section{Accelerating lattice gauge theory simulations}

\subsection{Why heat bath cannot be generalised}\label{sec:heat_bath}

The heat bath algorithm~\cite{PhysRevD.21.2308,Gattringer:2010zz} is an efficient way to sample new link variables locally in pure gauge theory. It effectively removes the potential rejection in standard Metropolis Hastings sampling. The heat bath uses the property that the gauge action $S_G$ is quadratic and diagonal in the plaquette $U_{\mu\nu}(n)$
\begin{align}
	S_G[U] &= \frac{\beta}{2N_c} \sum_{n,\mu<\nu}\tr \left[(1-U_{\mu\nu}(n))^\dagger(1-U_{\mu\nu}(n))\right]\,.
\end{align}
This allows to sample the plaquette very easily. So choosing a link, sampling all adjacent plaquettes while keeping the corresponding staples constant, and multiplying the plaquette by the inverse sum of staples, leads to an optimal update of the link.

Unfortunately, this updating scheme cannot be generalised to HMC type sampling because it operates in the group rather than the algebra. So, if we sample momenta corresponding to the group elements, the EOM will bring us out of the group. We could rescale the result to project it back into the group, but this changes the action, so most likely the update will be rejected. Heat bath only works because it does not require the link and corresponding plaquette to remain within the group over a continuous trajectory, only at discrete special time steps.

On a more abstract level, the problem is that we do not sample in flat space but on a manifold with non-trivial Haar measure which is not included in the action.

\subsection{Derivation of the approximative FA}\label{sec:qcd_derivations}

We write equation~\eqref{eq:link_as_exp} in the compact way
\begin{align}
	U_\mu(n) &= \eto{\im \Lambda_\mu(n)}\,,\\
	\Lambda_\mu(n) &= \frac12 x_{n,\mu}\cdot \lambda\,,
\end{align}
so that the plaquette can be expanded in small values of $\Lambda$
\begin{align}
	U_{\mu\nu}(n) &= U_\mu(n) U_\nu(n+\mu) U_\mu(n+\nu)^\dagger U_\nu(n)^\dagger\\
	&= \eto{\im\Lambda_\mu(n)} \eto{\im\Lambda_\nu(n+\mu)} \eto{-\im\Lambda_\mu(n+\nu)} \eto{-\im\Lambda_\nu(n)}\\
	\begin{split}		
		&= \left(1+\im\Lambda_\mu(n)-\frac12 \Lambda_\mu(n)^2\right)
		\left(1+\im\Lambda_\nu(n+\mu)-\frac12 \Lambda_\nu(n+\mu)^2\right)\\
		&\quad\times\left(1-\im\Lambda_\mu(n+\nu)-\frac12 \Lambda_\mu(n+\nu)^2\right)
		\left(1-\im\Lambda_\nu(n)-\frac12 \Lambda_\nu(n)^2\right)\\
		&\quad +\ordnung{\Lambda^3}\,.
	\end{split}
\end{align}
Under the (real part of the) trace required for the action only the even orders of $\Lambda$ survive
\begin{align} \begin{split}
	\Re\tr\left[1-U_{\mu\nu}(n)\right] &=
		\Re\tr\left[\vphantom{\frac12}\Lambda_\mu(n)\Lambda_\nu(n+\mu) - \Lambda_\mu(n)\Lambda_\mu(n+\nu)-\Lambda_\mu(n)\Lambda_\nu(n)\right.\\
		&\qquad -\Lambda_\nu(n+\mu)\Lambda_\mu(n+\nu) - \Lambda_\nu(n+\mu)\Lambda_\nu(n)+\Lambda_\mu(n+\nu)\Lambda_\nu(n)\\
		&\qquad\left.+\frac12\left(\Lambda_\mu(n)^2+\Lambda_\nu(n+\mu)^2+\Lambda_\mu(n+\nu)^2+\Lambda_\nu(n)^2\right)\right] + \ordnung{\Lambda^4}
	\end{split}\\
	\begin{split}
		&=\frac12\left[\vphantom{\frac12}x_{n,\mu}\cdot x_{n+\mu,\nu} - x_{n,\mu}\cdot x_{n+\nu,\mu}-x_{n,\mu}\cdot x_{n,\nu}\right.\\
		&\qquad -x_{n+\mu,\nu}\cdot x_{n+\nu,\mu} - x_{n+\mu,\nu}\cdot x_{n,\nu}+x_{n+\nu,\mu}\cdot x_{n,\nu}\\
		&\qquad\left.+\frac12\left(x_{n,\mu}^2+x_{n+\mu,\nu}^2+x_{n+\nu,\mu}^2+x_{n,\nu}^2\right)\right] + \ordnung{x^4}
	\end{split}
\end{align}
because all generators of SU$(N_c)$ are traceless (in case of U$(N_c)$ only the identity has non-zero trace and its odd power contributions are purely imaginary). In the step from $\Lambda$ to $x$ the generator orthogonality under the trace
\begin{align}
	\frac14\tr\left[\lambda_i\lambda_k\right] &= \frac 12 \delta_{ik}
\end{align}
was used.

Reintroducing the sum in the pure gauge action, allows to shift a number of indices, so that a Fourier transformation can be used to diagonalise the expression in real space
\begin{align}
	S_G(x) &= \frac\beta {N_c}\sum_{n,\mu<\nu}\Re\tr\left[1-U_{\mu\nu}(n)\right]\\
	&=  \frac\beta {2N_c}\sum_{n,\mu,\nu}\Re\tr\left[1-U_{\mu\nu}(n)\right]\\
	\begin{split}
		&=\frac{\beta}{4N_c}\sum_{n,\mu,\nu}\left[x_{n,\mu}\cdot x_{n+\mu,\nu} - x_{n,\mu}\cdot x_{n+\nu,\mu}-x_{n,\mu}\cdot x_{n,\nu}\right.\\
		&\qquad -x_{n,\mu}\cdot x_{n+\mu-\nu,\nu} - x_{n,\mu}\cdot x_{n-\nu,\mu}+x_{n,\mu}\cdot x_{n-\nu,\nu}\\
		&\qquad\left.+2 x_{n,\mu}^2\right] + \ordnung{x^4}
	\end{split}\\
	\begin{split}
		&=\frac{\beta}{4N_c}\sum_{k,\mu,\nu}\left[\eto{\im k_\mu}y_{k,\mu}^\dagger\cdot y_{k,\nu}^\pdagger -\eto{\im k_\nu} y_{k,\mu}^\dagger\cdot y_{k,\mu}^\pdagger-y_{k,\mu}^\dagger\cdot y_{k,\nu}^\pdagger\right.\\
		&\qquad -\eto{\im (k_\mu-k_\nu)}y_{k,\mu}^\dagger\cdot y_{k,\nu}^\pdagger - \eto{-\im k_\nu}y_{k,\mu}^\dagger\cdot y_{k,\mu}^\pdagger+\eto{-\im k_\mu}y_{k,\mu}^\dagger\cdot y_{k,\nu}^\pdagger\\
		&\qquad\left.+2 \left|y_{k,\mu}\right|^2\right] + \ordnung{y^4}\,,
	\end{split}\\
	y_{k,\mu} &= \frac{1}{\sqrt{N}}\sum_n \eto{-\im k\cdot n}x_{n,\mu}\,.
\end{align}
This result can be expressed more compactly with the help of some trigonometric identities
\begin{align}
	S_G(y) &= \frac\beta {N_c} \sum_{k,\mu,\nu} \left[\left|y_{k,\mu}\right|^2\sin\frac{k_\nu}{2} - y_{k,\mu}^\dagger\cdot y_{k,\nu}^\pdagger\, \eto{\im(k_\mu-k_\nu)}\sin\frac{k_\mu}{2}\sin\frac{k_\nu}{2}\right]\,,
\end{align}
obtaining the elegant form used in equation~\eqref{eq:m_of_k_pure_gauge}. Note that, induced by the dimensionality of $x$, each component of $y$ is a vector $y_{k,\mu}\in\mathbb{C}^{N_{\mathfrak{g}}}$ where $N_\mathfrak{g}$ is the number of generators $\lambda$, or equivalently the dimension of the corresponding Lie algebra. $M_G$ acts trivially within this space.

It is not straight forward to diagonalise the harmonic gauge matrix $M_G(k)$ analytically. However, the two distinct eigenvalues can be determined exactly and a (non-orthogonal) basis for each of the eigen-subspaces can be provided as follows. Let
\begin{align}
	y^0_k &= \sum_\mu \eto{\frac\im2 k_\mu}\sin\frac{k_\mu}{2}\,\hat{e}_\mu\,,\\
	y^i_k &= \eto{\frac\im2 k_\alpha}\sin\frac{k_i}{2}\,\hat{e}_\alpha - \eto{\frac\im2 k_i}\sin\frac{k_\alpha}{2}\,\hat{e}_i
\end{align}
with $\hat e_\mu$ the standard normal basis vectors. Here, in $d$ spatial dimensions, $i=1,\dots,d-1$ and $\alpha\in\{0,\dots,d-1\}$ is an arbitrary but fixed index so that $\sin\frac{k_\alpha}{2}\neq0$.
Then it is easy to check that the $y^0_k$, $y^i_k$ are eigenvectors of $M_G(k)$ with eigenvalues given by
\begin{align}
	M_G(k)\, y^0_k &= 0\,,\\
	M_G(k)\, y^i_k &= \left(\sum_\mu \sin^2\frac{k_\mu}{2} \right) y^i_k\,.
\end{align}
In practice, an efficient way to construct an orthonormal set of eigenvectors is to simply orthonormalise the raw eigenvectors $y^\mu_k$ using e.g.\ the modified Gram-Schmidt process.

This approximation works best in the weak coupling regime. Asymptotically as $\beta\rightarrow\infty$, it becomes exact and the FA based on this method is guaranteed to completely decorrelate the configurations after a single trajectory. For abelian groups the expansion is identical about every group element, not just the identity. Therefore, FA is guaranteed to work very well with abelian groups if the plaquette is close enough to the identity, even if the individual links are not. For non-abelian groups no such guarantee can be given as long as not all the links are close to the identity. In practice, however, this does not appear to be a major problem.

The optimal choice of the HMC Hamiltonian as by \cref{th:opt_traj_and_kin} is impossible because $M_G=\Omega_G\cdot \diag\left(\omega_G^2\right)\cdot\Omega_G^\dagger$ is not invertible. Instead, for this work we use
\begin{align}
	\mathcal{H} &= \frac12 p^\trans \tilde M_G^{-1} p + S_G(x)
\end{align}
with the regularised inverse
\begin{align}
	\tilde M_G^{-1} &\coloneqq \Omega_G\cdot \diag\left(\tilde\omega_G^{-2}\right)\cdot\Omega_G^\dagger\,,\\
	\tilde\omega_G^{-1} &\coloneqq \begin{cases}
		\omega_G^{-1} & \text{if } \omega_G\neq 0\,,\\
		\varepsilon_G & \text{if } \omega_G= 0\,,
	\end{cases}
\end{align}
where $\varepsilon_G=10^{-3}$ was chosen small without extensive tuning.

\subsection{Numerical verification of the approximate FA}\label{sec:lat_gauge_verify}

\begin{figure}[t]
	\centering
	\resizebox{0.98\textwidth}{!}{{\large%
\begingroup
  \inputencoding{latin1}%
  \makeatletter
  \providecommand\color[2][]{%
    \GenericError{(gnuplot) \space\space\space\@spaces}{%
      Package color not loaded in conjunction with
      terminal option `colourtext'%
    }{See the gnuplot documentation for explanation.%
    }{Either use 'blacktext' in gnuplot or load the package
      color.sty in LaTeX.}%
    \renewcommand\color[2][]{}%
  }%
  \providecommand\includegraphics[2][]{%
    \GenericError{(gnuplot) \space\space\space\@spaces}{%
      Package graphicx or graphics not loaded%
    }{See the gnuplot documentation for explanation.%
    }{The gnuplot epslatex terminal needs graphicx.sty or graphics.sty.}%
    \renewcommand\includegraphics[2][]{}%
  }%
  \providecommand\rotatebox[2]{#2}%
  \@ifundefined{ifGPcolor}{%
    \newif\ifGPcolor
    \GPcolortrue
  }{}%
  \@ifundefined{ifGPblacktext}{%
    \newif\ifGPblacktext
    \GPblacktexttrue
  }{}%
  \let\gplgaddtomacro\g@addto@macro
  \gdef\gplbacktext{}%
  \gdef\gplfronttext{}%
  \makeatother
  \ifGPblacktext
    \def\colorrgb#1{}%
    \def\colorgray#1{}%
  \else
    \ifGPcolor
      \def\colorrgb#1{\color[rgb]{#1}}%
      \def\colorgray#1{\color[gray]{#1}}%
      \expandafter\def\csname LTw\endcsname{\color{white}}%
      \expandafter\def\csname LTb\endcsname{\color{black}}%
      \expandafter\def\csname LTa\endcsname{\color{black}}%
      \expandafter\def\csname LT0\endcsname{\color[rgb]{1,0,0}}%
      \expandafter\def\csname LT1\endcsname{\color[rgb]{0,1,0}}%
      \expandafter\def\csname LT2\endcsname{\color[rgb]{0,0,1}}%
      \expandafter\def\csname LT3\endcsname{\color[rgb]{1,0,1}}%
      \expandafter\def\csname LT4\endcsname{\color[rgb]{0,1,1}}%
      \expandafter\def\csname LT5\endcsname{\color[rgb]{1,1,0}}%
      \expandafter\def\csname LT6\endcsname{\color[rgb]{0,0,0}}%
      \expandafter\def\csname LT7\endcsname{\color[rgb]{1,0.3,0}}%
      \expandafter\def\csname LT8\endcsname{\color[rgb]{0.5,0.5,0.5}}%
    \else
      \def\colorrgb#1{\color{black}}%
      \def\colorgray#1{\color[gray]{#1}}%
      \expandafter\def\csname LTw\endcsname{\color{white}}%
      \expandafter\def\csname LTb\endcsname{\color{black}}%
      \expandafter\def\csname LTa\endcsname{\color{black}}%
      \expandafter\def\csname LT0\endcsname{\color{black}}%
      \expandafter\def\csname LT1\endcsname{\color{black}}%
      \expandafter\def\csname LT2\endcsname{\color{black}}%
      \expandafter\def\csname LT3\endcsname{\color{black}}%
      \expandafter\def\csname LT4\endcsname{\color{black}}%
      \expandafter\def\csname LT5\endcsname{\color{black}}%
      \expandafter\def\csname LT6\endcsname{\color{black}}%
      \expandafter\def\csname LT7\endcsname{\color{black}}%
      \expandafter\def\csname LT8\endcsname{\color{black}}%
    \fi
  \fi
    \setlength{\unitlength}{0.0500bp}%
    \ifx\gptboxheight\undefined%
      \newlength{\gptboxheight}%
      \newlength{\gptboxwidth}%
      \newsavebox{\gptboxtext}%
    \fi%
    \setlength{\fboxrule}{0.5pt}%
    \setlength{\fboxsep}{1pt}%
    \definecolor{tbcol}{rgb}{1,1,1}%
\begin{picture}(7200.00,5040.00)%
    \gplgaddtomacro\gplbacktext{%
      \csname LTb\endcsname%
      \put(814,704){\makebox(0,0)[r]{\strut{}$0$}}%
      \csname LTb\endcsname%
      \put(814,1527){\makebox(0,0)[r]{\strut{}$0.2$}}%
      \csname LTb\endcsname%
      \put(814,2350){\makebox(0,0)[r]{\strut{}$0.4$}}%
      \csname LTb\endcsname%
      \put(814,3173){\makebox(0,0)[r]{\strut{}$0.6$}}%
      \csname LTb\endcsname%
      \put(814,3996){\makebox(0,0)[r]{\strut{}$0.8$}}%
      \csname LTb\endcsname%
      \put(814,4819){\makebox(0,0)[r]{\strut{}$1$}}%
      \csname LTb\endcsname%
      \put(946,484){\makebox(0,0){\strut{}$0$}}%
      \csname LTb\endcsname%
      \put(1597,484){\makebox(0,0){\strut{}$2$}}%
      \csname LTb\endcsname%
      \put(2248,484){\makebox(0,0){\strut{}$4$}}%
      \csname LTb\endcsname%
      \put(2898,484){\makebox(0,0){\strut{}$6$}}%
      \csname LTb\endcsname%
      \put(3549,484){\makebox(0,0){\strut{}$8$}}%
      \csname LTb\endcsname%
      \put(4200,484){\makebox(0,0){\strut{}$10$}}%
      \csname LTb\endcsname%
      \put(4851,484){\makebox(0,0){\strut{}$12$}}%
      \csname LTb\endcsname%
      \put(5501,484){\makebox(0,0){\strut{}$14$}}%
      \csname LTb\endcsname%
      \put(6152,484){\makebox(0,0){\strut{}$16$}}%
      \csname LTb\endcsname%
      \put(6803,484){\makebox(0,0){\strut{}$18$}}%
    }%
    \gplgaddtomacro\gplfronttext{%
      \csname LTb\endcsname%
      \put(5816,1922){\makebox(0,0)[r]{\strut{}FA, $T=\nicefrac\pi2$}}%
      \csname LTb\endcsname%
      \put(5816,1592){\makebox(0,0)[r]{\strut{}no FA, $T=\nicefrac{1}{\sqrt{\beta}}$}}%
      \csname LTb\endcsname%
      \put(5816,1262){\makebox(0,0)[r]{\strut{}strong coupling expansion}}%
      \csname LTb\endcsname%
      \put(5816,932){\makebox(0,0)[r]{\strut{}weak coupling expansion}}%
      \csname LTb\endcsname%
      \put(209,2761){\rotatebox{-270.00}{\makebox(0,0){\strut{}$\erwartung{P(\beta)}$}}}%
      \put(3874,154){\makebox(0,0){\strut{}$\beta$}}%
    }%
    \gplbacktext
    \put(0,0){\includegraphics[width={360.00bp},height={252.00bp}]{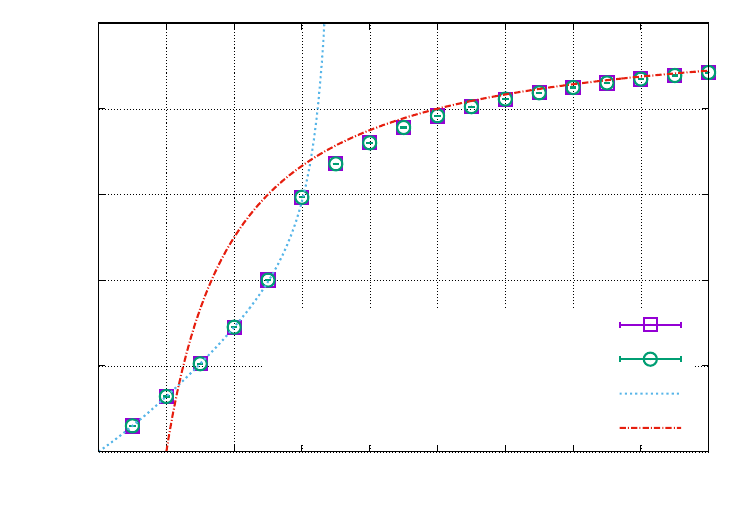}}%
    \gplfronttext
  \end{picture}%
\endgroup
\begingroup
  \inputencoding{latin1}%
  \makeatletter
  \providecommand\color[2][]{%
    \GenericError{(gnuplot) \space\space\space\@spaces}{%
      Package color not loaded in conjunction with
      terminal option `colourtext'%
    }{See the gnuplot documentation for explanation.%
    }{Either use 'blacktext' in gnuplot or load the package
      color.sty in LaTeX.}%
    \renewcommand\color[2][]{}%
  }%
  \providecommand\includegraphics[2][]{%
    \GenericError{(gnuplot) \space\space\space\@spaces}{%
      Package graphicx or graphics not loaded%
    }{See the gnuplot documentation for explanation.%
    }{The gnuplot epslatex terminal needs graphicx.sty or graphics.sty.}%
    \renewcommand\includegraphics[2][]{}%
  }%
  \providecommand\rotatebox[2]{#2}%
  \@ifundefined{ifGPcolor}{%
    \newif\ifGPcolor
    \GPcolortrue
  }{}%
  \@ifundefined{ifGPblacktext}{%
    \newif\ifGPblacktext
    \GPblacktexttrue
  }{}%
  \let\gplgaddtomacro\g@addto@macro
  \gdef\gplbacktext{}%
  \gdef\gplfronttext{}%
  \makeatother
  \ifGPblacktext
    \def\colorrgb#1{}%
    \def\colorgray#1{}%
  \else
    \ifGPcolor
      \def\colorrgb#1{\color[rgb]{#1}}%
      \def\colorgray#1{\color[gray]{#1}}%
      \expandafter\def\csname LTw\endcsname{\color{white}}%
      \expandafter\def\csname LTb\endcsname{\color{black}}%
      \expandafter\def\csname LTa\endcsname{\color{black}}%
      \expandafter\def\csname LT0\endcsname{\color[rgb]{1,0,0}}%
      \expandafter\def\csname LT1\endcsname{\color[rgb]{0,1,0}}%
      \expandafter\def\csname LT2\endcsname{\color[rgb]{0,0,1}}%
      \expandafter\def\csname LT3\endcsname{\color[rgb]{1,0,1}}%
      \expandafter\def\csname LT4\endcsname{\color[rgb]{0,1,1}}%
      \expandafter\def\csname LT5\endcsname{\color[rgb]{1,1,0}}%
      \expandafter\def\csname LT6\endcsname{\color[rgb]{0,0,0}}%
      \expandafter\def\csname LT7\endcsname{\color[rgb]{1,0.3,0}}%
      \expandafter\def\csname LT8\endcsname{\color[rgb]{0.5,0.5,0.5}}%
    \else
      \def\colorrgb#1{\color{black}}%
      \def\colorgray#1{\color[gray]{#1}}%
      \expandafter\def\csname LTw\endcsname{\color{white}}%
      \expandafter\def\csname LTb\endcsname{\color{black}}%
      \expandafter\def\csname LTa\endcsname{\color{black}}%
      \expandafter\def\csname LT0\endcsname{\color{black}}%
      \expandafter\def\csname LT1\endcsname{\color{black}}%
      \expandafter\def\csname LT2\endcsname{\color{black}}%
      \expandafter\def\csname LT3\endcsname{\color{black}}%
      \expandafter\def\csname LT4\endcsname{\color{black}}%
      \expandafter\def\csname LT5\endcsname{\color{black}}%
      \expandafter\def\csname LT6\endcsname{\color{black}}%
      \expandafter\def\csname LT7\endcsname{\color{black}}%
      \expandafter\def\csname LT8\endcsname{\color{black}}%
    \fi
  \fi
    \setlength{\unitlength}{0.0500bp}%
    \ifx\gptboxheight\undefined%
      \newlength{\gptboxheight}%
      \newlength{\gptboxwidth}%
      \newsavebox{\gptboxtext}%
    \fi%
    \setlength{\fboxrule}{0.5pt}%
    \setlength{\fboxsep}{1pt}%
    \definecolor{tbcol}{rgb}{1,1,1}%
\begin{picture}(7200.00,5040.00)%
    \gplgaddtomacro\gplbacktext{%
      \csname LTb\endcsname%
      \put(946,704){\makebox(0,0)[r]{\strut{}$10^{-6}$}}%
      \csname LTb\endcsname%
      \put(946,1733){\makebox(0,0)[r]{\strut{}$10^{-5}$}}%
      \csname LTb\endcsname%
      \put(946,2762){\makebox(0,0)[r]{\strut{}$10^{-4}$}}%
      \csname LTb\endcsname%
      \put(946,3790){\makebox(0,0)[r]{\strut{}$10^{-3}$}}%
      \csname LTb\endcsname%
      \put(946,4819){\makebox(0,0)[r]{\strut{}$10^{-2}$}}%
      \csname LTb\endcsname%
      \put(1078,484){\makebox(0,0){\strut{}$0$}}%
      \csname LTb\endcsname%
      \put(2032,484){\makebox(0,0){\strut{}$1$}}%
      \csname LTb\endcsname%
      \put(2986,484){\makebox(0,0){\strut{}$2$}}%
      \csname LTb\endcsname%
      \put(3941,484){\makebox(0,0){\strut{}$3$}}%
      \csname LTb\endcsname%
      \put(4895,484){\makebox(0,0){\strut{}$4$}}%
      \csname LTb\endcsname%
      \put(5849,484){\makebox(0,0){\strut{}$5$}}%
      \csname LTb\endcsname%
      \put(6803,484){\makebox(0,0){\strut{}$6$}}%
    }%
    \gplgaddtomacro\gplfronttext{%
      \csname LTb\endcsname%
      \put(4371,4591){\makebox(0,0)[r]{\strut{}FA, $T=\nicefrac\pi2$}}%
      \csname LTb\endcsname%
      \put(4371,4261){\makebox(0,0)[r]{\strut{}no FA, $T=\nicefrac{1}{\sqrt{\beta}}$}}%
      \csname LTb\endcsname%
      \put(209,2761){\rotatebox{-270.00}{\makebox(0,0){\strut{}$\left|\Delta\erwartung{P(\beta)}\right|$}}}%
      \put(3940,154){\makebox(0,0){\strut{}$\beta$}}%
    }%
    \gplbacktext
    \put(0,0){\includegraphics[width={360.00bp},height={252.00bp}]{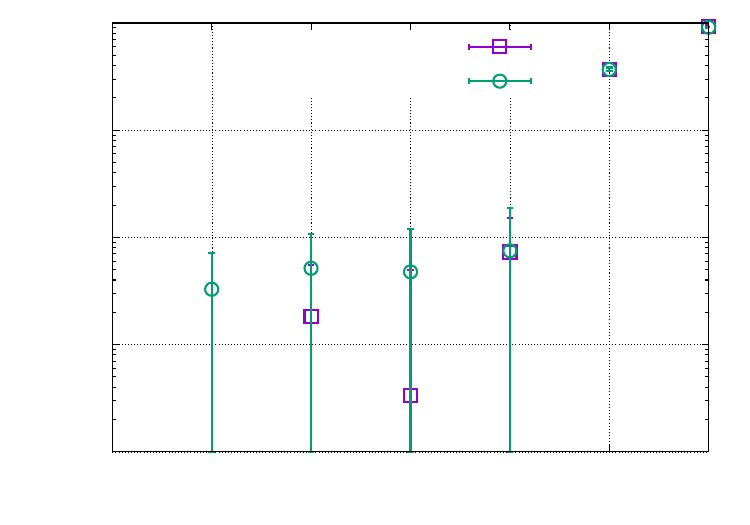}}%
    \gplfronttext
  \end{picture}%
\endgroup
}}
	\caption{Plaquette results in SU$(3)$ pure gauge theory on a $10^4$ lattice from HMC simulations checked against the strong coupling expansion~\cite{PhysRevD.11.2104,PhysRevD.19.2514} using eq.~\eqref{eq:strong_coupling_exp} (the 16th order expansion was converted to Padé approximation form in order to maximise the range of validity) and the weak coupling expansion~\cite{Creutz:1983njd} using eq.~\eqref{eq:weak_coupling_exp}. Left: plaquette expectation value $\erwartung{P(\beta)}$; Right: absolute value of the deviation of $\erwartung{P(\beta)}$ from eq.~\eqref{eq:strong_coupling_exp}.}
	\label{fig:pure_gauge_check}
\end{figure}

We benchmark the results of our numerical HMC simulations with and without FA against the strong coupling expansion in figure~\ref{fig:pure_gauge_check}. Refs.~\cite{PhysRevD.11.2104,PhysRevD.19.2514} provide explicit formulae for the free energy $F$ up to 16th order. The free energy can be translated to the plaquette expectation value as
\begin{align}
	\erwartung{P(\beta)} &\equiv \erwartung{\frac{1}{L^d}\sum_{n}\frac{2}{d(d-1)}\sum_{\mu<\nu}\frac{1}{N_c}\Re\tr\left[U_{\mu\nu}(n)\right]}\\
	&= \frac{2}{d(d-1)\,N_c!}\,\frac{\md}{\md \beta} F\left(\frac{\beta}{N_c!}\right)\,.\label{eq:strong_coupling_exp}
\end{align}
For large $\beta$ the weak coupling expansion~\cite{Creutz:1983njd}
\begin{align}
	\erwartung{P(\beta)} &= 1-2/\beta+\ordnung{\beta^{-2}}\label{eq:weak_coupling_exp}
\end{align}
becomes applicable.
As expected, we find excellent agreement between (a) both numerical methods and the strong coupling expansion for $\beta\lesssim 4$, (b) both numerical methods and the weak coupling expansion for $\beta\gtrsim 14$, and (c) the two numerical methods with each other for the entire tested parameter range.

\subsection{Alternative approach sampling the plaquette}\label{sec:plaquette_sampling}

From \cref{sec:heat_bath} we learn that we have to sample in the algebra, as is done in classical lattice QCD HMC simulations for exactly this reason. One alternative option to sampling the links as in \cref{sec:qcd_derivations} is to still sample the plaquette like heat bath does. This means that not all the links would be updated at the same time. Instead, for every update step, one would have to choose a subset of links to update while keeping the rest fixed. Such a subset can be extensive in the volume, so global updates would still be possible. Expand the plaquette $U_{\mu\nu}(n)=\exp\left(\im X_{n,\mu\nu}\right)$ around unity to obtain $S_G \sim \beta X^2$. Now sample momenta according to this harmonic part and then evolve the EOM using the force from the full action. Finally update the links within the chosen subset using the changed plaquettes and multiplying by the inverse sum of fixed staples (same as heat bath).

If $\beta$ is large, then deviations from unity will be strongly suppressed, so the harmonic part can be solved exactly (i.e.\ with EFA) and the deviation of the gauge action from the harmonic approximation can be treated as part of the perturbation. If $\beta$ is not very large, then the approximation is not very good either and it makes more sense to use the full action for updates directly and only make discrete time steps. In any case, sampling momenta according to this distribution should be close to optimal for the decorrelation.

The advantage of this ansatz over the link sampling in \cref{sec:qcd_derivations} is that the $U_\mu(n)$ are not required to be close to the identity link by link. It suffices that the plaquette is close to the identity. In addition, no Fourier transformation is involved. On the other hand $\ordnung{d^2}$ updates are required for a full update of all links, as opposed to a single global update in \cref{sec:qcd_derivations}. This overhead might be especially crucial when expensive fermionic updates are involved.

Overall, this method and the link sampling both appear viable. Potentially, the plaquette sampling is advantageous at moderately weak coupling where the plaquettes are close to unity but the links are not. \end{document}